%% file: BODY.tex
\documentclass{article} 
\usepackage{iclr2027_conference,times}

\input{math_commands.tex}

\usepackage{hyperref}
\usepackage{url}

\usepackage{booktabs}
\usepackage{graphicx}
\usepackage{kotex}
\usepackage{multirow}
\usepackage[table]{xcolor}
\usepackage{array}
\usepackage{wrapfig}
\usepackage{arydshln}
\usepackage{caption}
\usepackage{colortbl}
\usepackage[export]{adjustbox}
\usepackage{tabularx}

\title{RenderRank: Learning to Rerank Text\\ with Compressed Visual Tokens}

\author{Seongtae Hong\textsuperscript{1}, ~~Youngjoon Jang\textsuperscript{1}, ~~Jungseob Lee\textsuperscript{1}, ~~Hyeonseok Moon\textsuperscript{2}, ~~Heuiseok Lim\textsuperscript{1}\thanks{Corresponding author} \\
\textsuperscript{1}Korea University \\ 
\textsuperscript{2}Sookmyung Women's University \\[3pt]
\texttt{ghdchlwls123@korea.ac.kr}
}

\iclrfinalcopy 
\begin{document}

\maketitle

\begin{abstract}
Rendering document text as images allows vision-language models to encode documents as visual tokens, which can reduce input sequence length compared with text input. This reduction in input length is particularly useful for reranking, where each query involves scoring multiple candidate documents and token savings apply to each candidate evaluation. We introduce RenderRank, a reranker that learns query-dependent relevance scoring from compressed visual document representations instead of the text token sequences used by conventional text-based rerankers. Training first aligns relevance scores from visual inputs with those of a text-based teacher, then refines the relative scores of positive and negative documents for the same query. Across 11 datasets from BEIR, RenderRank uses 16.5--35.5\% fewer input tokens while achieving an average NDCG@10 of 55.96, outperforming all evaluated text-based baselines below 4B parameters and some larger models. Across four long-document datasets, it achieves an average NDCG@10 of 88.27 with approximately half the average input token count of the evaluated text-based rerankers. In this setting, RenderRank delivers $1.70\times$ the highest average throughput of the evaluated baselines. These results demonstrate that compressed visual representations can support accurate document relevance scoring, providing an alternative to text token representations for reranking.
\end{abstract}

\input{Texs/Introduction}

\input{Texs/Related_Work}
\input{Texs/Methods}
\input{Texs/Experiments}

\input{Texs/Results}
\input{Texs/Analysis_Ablation}

\input{Texs/Conclusion}


\bibliography{iclr2027_conference}
\bibliographystyle{iclr2027_conference}

\appendix
\input{Texs/Appendix}

\end{document}

%% file: math_commands.tex
\usepackage{amsmath,amsfonts,bm}

\def\eqref#1{equation~\ref{#1}}

\def\1{\bm{1}}

\DeclareMathAlphabet{\mathsfit}{\encodingdefault}{\sfdefault}{m}{sl}
\SetMathAlphabet{\mathsfit}{bold}{\encodingdefault}{\sfdefault}{bx}{n}



%% file: Texs/Introduction.tex
\section{Introduction}
\label{sec:introduction}
Retrieval-augmented generation (RAG) generates answers to queries using information retrieved from external documents as supporting evidence~\citep{lewis2020retrieval,izacard2021leveraging,izacard2023atlas}. In this pipeline, first-stage retrieval efficiently identifies potentially relevant candidate documents from a large corpus~\citep{karpukhin2020dense,xiong2020approximate,khattab2020colbert}. However, retrieved candidates vary in their relevance to the query and their usefulness as evidence for answering it, so a more precise assessment is needed to determine which documents to use for subsequent generation~\citep{yu2024rankrag,asai2024self}. A reranker prioritizes candidate documents for generation based on their relevance to the query~\citep{nogueira2019passage,glass2022re2g}. Reranking computes a relevance score for each retrieved candidate, with the query and the corresponding document provided together as input~\citep{nogueira2020document,qwen3embedding}.

When reranking multiple candidates, the input length of each document affects the computational cost of evaluating the entire candidate set. Longer token sequences require more computation for relevance scoring, and these costs accumulate across retrieved candidates~\citep{peng-etal-2025-efficiency}. Limiting the input length can reduce reranking computation, but may exclude evidence or context needed to assess relevance to the query~\citep{li2023parade}. This highlights the need to represent document content with fewer tokens while preserving the information needed for relevance assessment. Such an approach could enable relevance scoring for the same document with fewer tokens, or allow more document content to be considered within the same sequence length limit.

From this perspective, rendering document text as images and encoding them into visual tokens offers a potential way to shorten document representations. Instead of feeding the source text directly as text tokens, this approach uses the visual encoder of a vision-language model (VLM) to convert the rendered images into visual tokens for the language backbone~\citep{wang2024qwen2}. In generation tasks, such visual representations have been used to perform question answering and summarization with fewer input tokens~\citep{li2025text}. Visual text compression has further enabled models to process more source content within a limited context window and reduce the time spent on input processing and answer generation~\citep{cheng2026glyph}. These results show that models can understand and use document content for downstream tasks even when it is represented as a shorter sequence of visual tokens.

Extending the benefits of visual text compression to reranking requires the ability to assess relevance to a query from compressed document content. Beyond changing the document input modality from text to images, the rendered images must enable the model to understand document content. Rendering configurations affect both the length of the visual token sequence and how readily the model can interpret the content. Shorter sequences can reduce attention and token-wise computation in the language model, accelerating relevance scoring even when each candidate is scored in a single forward pass. These configurations must therefore account for both computational efficiency and document understanding. Alongside these configurations, training must adapt relevance scoring to visual inputs by teaching the model to relate these visual representations to the query.

\begin{figure*}[t!]
    \centering
    \includegraphics[width=0.85\textwidth]{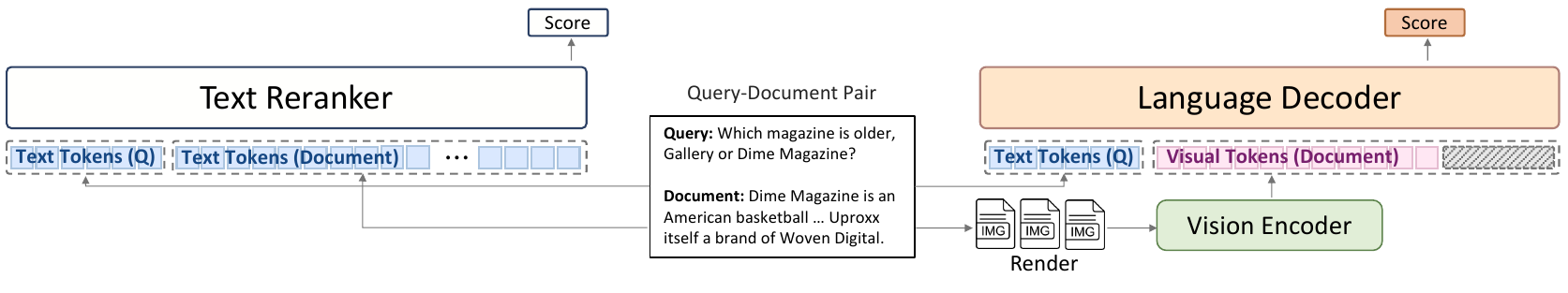}
    \caption{Overview of conventional text-based reranking (left) and RenderRank (right) for the same query--document pair. RenderRank encodes rendered document images into visual tokens, which the language decoder processes with the textual query to compute a relevance score.}
    \label{fig:struc}
\end{figure*}

We introduce RenderRank, a reranker that renders document text as images, encodes them into compressed visual tokens, and scores document relevance to a textual query. Figure~\ref{fig:struc} provides an overview of RenderRank and illustrates how its document representation and relevance scoring pipeline differ from those of a text-based reranker. We employ a two-stage training approach to learn relevance scoring from visual document representations and to assign higher scores to positive documents than to negatives. First, Cross-Modal Relevance Distillation trains the model on textual queries and rendered document images to predict relevance scores that approximate those assigned by a teacher to the corresponding textual inputs. Next, Query-Local Relevance Discrimination refines relevance scoring by learning the relative scores of positive and negative documents associated with the same query. Document images are encoded in advance, independently of the query, and the resulting visual representations are used to predict relevance scores conditioned on the textual query.

RenderRank is evaluated on 11 datasets from BEIR~\citep{beir} and four long-document datasets in terms of ranking quality, input token counts, estimated computational cost, and measured throughput. RenderRank achieves an average NDCG@10 of 55.96 on BEIR, outperforming all other evaluated models with fewer than 4B parameters and some larger models. It uses 16.5--35.5\% fewer input tokens than the text-based rerankers evaluated. RenderRank also achieves higher inference throughput than models of similar size and some smaller models, showing that its efficiency benefits extend beyond input token savings. For long-document reranking, it achieves high reranking performance with approximately half the average input token count of the text-based rerankers evaluated and outperforms all compared models at every tested maximum sequence length under identical length limits. These results demonstrate that representing document text as visual tokens supports high reranking effectiveness and inference efficiency while enabling more document content to be used within a limited input length.

%% file: Texs/Related_Work.tex
\section{Related Work}
\label{sec:related_work}

\paragraph{Visual Text Representation and Compression}
Rendering text as images allows document content to be represented through visual tokens rather than text
tokens~\citep{rust2022language,lyu2025pixelworld,tschannen2023clippo,xiao2024pixel}. Beyond understanding existing document images, this approach considers how source text should be visually structured for model input~\citep{lotz2023text}. Prior work explores both the feasibility of understanding text through visual representations and the potential to reduce inference cost and latency by using fewer tokens to represent documents~\citep{li2025text,cheng2026glyph}. Visual information preservation has also been examined in relation to rendering choices such as font size, line spacing, and page layout~\citep{tang2026visual}. Rendering density affects both the preservation of fine-grained textual information and the amount of source text represented by each visual token, allowing fewer tokens to encode a document or more content to fit within the same sequence length limit. Complementary approaches diversify rendering during training to reduce reliance on superficial visual cues~\citep{yuan2026design}.

\paragraph{Document Reranking}
Reranking reorders candidate documents returned by a retriever according to their relevance to a query. Cross-encoder rerankers score relevance by modeling interactions between the query and document~\citep{nogueira2019passage,nogueira2020document}. Large language models have been adopted for reranking to leverage their language understanding and reasoning capabilities for improved effectiveness~\citep{sun2023chatgpt,ma2024fine,qwen3embedding}. Building on their reranking performance, knowledge distillation approaches use these models as teachers, training other rerankers on the relative ordering of candidate documents~\citep{baldelli2024twolar,schlatt2025rank}. Distillation objectives also include directly matching the teacher's relevance score differences between documents~\citep{hofstatter2020improving}. As inference costs increase with model size, document length, and the number of candidates evaluated per query, reranking effectiveness is also studied alongside computational cost and throughput~\citep{peng-etal-2025-efficiency,aarsen2026ettin-reranker}. In image–text reranking, precomputing visual features and compressing visual tokens have been explored to reduce the computational cost of scoring candidate pairs~\citep{taraday2025efficient}. Efficient reranking of document page images has also been studied through pretraining on rendered text followed by further training on document images, using query-dependent visual token selection and a listwise approach that ranks the candidate set in a single forward pass~\citep{sun2026very}.

RenderRank focuses on reducing the input sequence length required for reranking by encoding candidate documents originally provided as text into compressed visual tokens. To this end, it learns to approximate relevance scores from a text-based teacher using visual inputs, then refines the relative scores of candidates associated with the same query. By representing document content with fewer tokens, this approach allows longer documents to fit within the same sequence length limit for relevance scoring.

%% file: Texs/Methods.tex
\section{RenderRank}
\label{sec:RenderRank}
RenderRank renders document text as images and encodes them into visual tokens to score relevance to a textual query. To this end, we render documents with consideration for text legibility and information density, and train the model to assess relevance between visual document representations and textual queries. This section describes the document rendering procedure and the rationale for its configuration, followed by the reranking training method based on these visual representations.

\input{Tables/Generation}
\subsection{Rendering Text for Visual Processing}

Rendering document text as images requires representing sufficient content within a limited image size while preserving the character shapes and layout needed for the model to interpret the text. Smaller fonts and tighter line spacing can increase information density and reduce the number of visual tokens, but fewer pixels per character and less separation between lines may make the content harder to recognize. Rendering configurations should therefore account for both token reduction and document understanding. We use question answering and summarization to evaluate how well models understand and use rendered document content. Question answering evaluates the ability to identify and use evidence relevant to a query, whereas summarization evaluates the ability to identify and synthesize the main content of a document. We therefore use performance on these two tasks and token efficiency to determine the rendering configuration for reranking, while considering the potential reuse of rendered documents as inputs to a downstream generation model. Specifically, we adopt Roboto Regular as used by \citet{tang2026visual} and vary font size and line spacing to compare generation performance and input token counts.

Table~\ref{tab:rendering_config} compares question answering and summarization performance and token reduction rates across font sizes and line spacings. With line spacing fixed at 1.0, 12pt yields less token reduction than 10pt but higher performance on both tasks for both models. For example, the QASPER F1 score of Qwen3.6 increases from 41.32 to 48.72. In contrast, increasing line spacing at 12pt increases the number of visual tokens without consistent performance gains. These results suggest that an appropriate rendering configuration can reduce token usage while largely preserving task performance.

\paragraph{Rendering Configuration}
Based on the performance and token efficiency observed in the preceding experiments, we render document text for training RenderRank in Roboto Regular at 12pt with line spacing of 1.0. Image width and height are set to multiples of 32 pixels to accommodate the $16\times16$-pixel patches and $2\times2$ spatial merging of the Qwen3-VL~\citep{qwen3vlembedding} visual encoder used in RenderRank. Documents are rendered at 96 DPI with a fixed width of 896 pixels and a maximum height of 896 pixels, with the height adjusted in 32-pixel increments according to the content. Text is rendered in black on a white background without margins and wraps to the next line when it would exceed the image width. Content exceeding one page continues on the next image without overlap. Long documents are rendered into multiple images, which preserve the order of the document content and are processed as a single candidate document.

\subsection{Learning to Rerank Text through Visual Representations}
RenderRank is trained to assess document relevance by relating rendered document images to a textual query. Let $q$ denote the query, $d$ the original document, and $R(d)$ the sequence of rendered document images. Parameterized by $\theta$, RenderRank takes the textual query and the document's visual tokens as input and produces a relevance score $s_{\theta}(q,R(d))$. Training proceeds in two stages: (1) Cross-Modal Relevance Distillation, which trains the model to predict a text-based teacher's relevance scores from visual inputs, and (2) Query-Local Relevance Discrimination, which refines relevance scoring to prioritize the positive document within the candidate set for each query.

\paragraph{Cross-Modal Relevance Distillation}

Using visual tokens to represent document text for reranking requires the ability to relate the content of document images to a query and assess its relevance. However, changing the input representation alone does not ensure that a text-based reranker's relevance scoring capabilities are preserved with visual inputs. We therefore perform cross-modal relevance distillation by pairing the text and image representations of the same document. The student learns to approximate a text-based teacher's relevance scores from the textual query and rendered document images.

Let $t(q,d)$ denote the relevance score produced by the teacher from the textual query $q$ and document $d$, and $s_{\theta}(q,R(d))$ the score produced by the student from the same query and rendered document images $R(d)$. We perform cross-modal relevance distillation by minimizing the mean squared error (MSE) between these scores over the set of training pairs $\mathcal{D}$:
\begin{equation}
\label{eq:cross_modal_relevance_distillation}
\mathcal{L}_{\mathrm{CMRD}}
=
\mathbb{E}_{(q,d)\sim\mathcal{D}}
\left[
\bigl(s_{\theta}(q,R(d))-t(q,d)\bigr)^2
\right]
\end{equation}
This supervision encourages cross-modal alignment at the level of relevance scores without explicitly aligning text and image features in a shared representation space. Conditioned on the same query, the student learns to assign the rendered document a score consistent with the teacher's score for its textual counterpart. By operating at the score level, this objective supports distillation between textual and visual sequences of different lengths without requiring token-level correspondence.

\paragraph{Query-Local Relevance Discrimination}

Building on the cross-modal score alignment learned in the preceding stage, we train the model to assign higher scores to the positive document than to negatives within each query's candidate set, refining visual-input relevance prediction for reranking. For each query $q_i$, we construct a candidate set $\mathcal{C}_{i}=\{d_i^{+},d_{i,1}^{-},\ldots,d_{i,K}^{-}\}$, where $d_i^{+}$ is the positive document and $\{d_{i,j}^{-}\}_{j=1}^{K}$ are the $K$ negative documents for the same query. We use an InfoNCE contrastive loss restricted to each query's candidate set, with the objective defined as follows:
\begin{equation}
\label{eq:query_local_relevance_discrimination}
    \mathcal{L}_{\mathrm{QLRD}}
    =
    -\frac{1}{N}\sum_{i=1}^{N}
    \log
    \frac{
    \exp\bigl(s_{\theta}(q_i,R(d_i^{+}))\bigr)
    }{
    \exp\bigl(s_{\theta}(q_i,R(d_i^{+}))\bigr)
    +
    \sum_{j=1}^{K}
    \exp\bigl(s_{\theta}(q_i,R(d_{i,j}^{-}))\bigr)
    }
\end{equation}
Here, $N$ denotes the number of queries in a mini-batch. This objective trains the model to assess the relevance of rendered document content through contrasts between positive and negative documents associated with the same query. Although the loss compares candidates within each query, the model computes each relevance score using only the query and the corresponding document.

%% file: Tables/Generation.tex
\definecolor{renderblue}{HTML}{2196D2}
\definecolor{renderred}{HTML}{E57373}

\begin{wraptable}{r}{0.35\textwidth}
\vspace{-13pt}
  \centering
  \caption{Rendering configurations evaluated by F1 on QASPER and ROUGE-L on GovReport. Blue and red indicate fewer and more tokens than text input, respectively.}
  \label{tab:rendering_config}
  \setlength{\tabcolsep}{4pt}
  \renewcommand{\arraystretch}{1.0}
  \resizebox{\linewidth}{!}{%
  \begin{tabular}{cccccc}
  \toprule
  \multirow{2}{*}{Font}
  & \multirow{2}{*}{LS}
  & \multicolumn{2}{c}{Gemma-4-26B-A4B-it}
  & \multicolumn{2}{c}{Qwen3.6-35B-A3B} \\
  \cmidrule(lr){3-4}\cmidrule(lr){5-6}
  & & QASPER & GovReport & QASPER & GovReport \\
    \midrule
    Text & --
    & 47.73 & 28.78 & 51.45 & 31.10 \\
    \midrule
    \multirow{3}{*}{10pt} & 1.0
    & \cellcolor{renderblue!19.35}45.90
    & \cellcolor{renderblue!20.51}28.27
    & \cellcolor{renderblue!30.98}41.32
    & \cellcolor{renderblue!30.16}30.55 \\
    & 1.1
    & \cellcolor{renderblue!15.65}45.72
    & \cellcolor{renderblue!16.40}28.46
    & \cellcolor{renderblue!28.18}43.18
    & \cellcolor{renderblue!27.20}30.75 \\
    & 1.2
    & \cellcolor{renderblue!14.33}44.58
    & \cellcolor{renderblue!14.70}28.27
    & \cellcolor{renderblue!27.08}45.80
    & \cellcolor{renderblue!25.99}30.42 \\
    \hline
    \multirow{3}{*}{11pt} & 1.0
    & \cellcolor{renderblue!11.32}47.48
    & \cellcolor{renderblue!11.53}28.53
    & \cellcolor{renderblue!24.84}45.26
    & \cellcolor{renderblue!23.67}30.53 \\
    & 1.1
    & \cellcolor{renderblue!6.38}48.32
    & \cellcolor{renderblue!6.76}28.69
    & \cellcolor{renderblue!21.67}46.46
    & \cellcolor{renderblue!20.21}30.55 \\
    & 1.2
    & \cellcolor{renderblue!4.19}45.76
    & \cellcolor{renderblue!4.05}28.68
    & \cellcolor{renderblue!19.95}47.41
    & \cellcolor{renderblue!18.44}30.85 \\
    \hline
    \multirow{3}{*}{12pt} & 1.0
    & \cellcolor{renderblue!6.59}48.07
    & \cellcolor{renderblue!7.01}28.67
    & \cellcolor{renderblue!21.83}48.72
    & \cellcolor{renderblue!20.40}30.82 \\
    & 1.1
    & \cellcolor{renderblue!1.50}47.93
    & \cellcolor{renderblue!1.35}28.75
    & \cellcolor{renderblue!17.91}45.35
    & \cellcolor{renderblue!16.27}30.66 \\
    & 1.2
    & \cellcolor{renderred!17.76}46.50
    & \cellcolor{renderred!19.03}28.60
    & \cellcolor{renderblue!14.59}45.33
    & \cellcolor{renderblue!12.46}30.69 \\
    \bottomrule
  \end{tabular}%
  }
  \vspace{-30pt}
  \end{wraptable}

%% file: Texs/Experiments.tex
\section{Experimental Setup}
\label{sec:expsestup}
\paragraph{Training}
RenderRank is initialized from Qwen3-VL-Reranker-2B~\citep{qwen3vlembedding} and trained in two stages. For cross-modal relevance distillation, we use the English fine-tuning data released by \citet{sourty2026denseonlateonfullyopen}, comprising 1.57M training records with one positive document and ten negatives per query. We retain all records and select one positive and three negatives per query, yielding 6.28M query--document pairs. We use Qwen3-Reranker-4B~\citep{qwen3embedding} as the teacher model to provide relevance scores for the original text query--document pairs. 

For query-local relevance discrimination, we use RLHN-100K~\citep{rlhn}, with the positive and negative documents associated with each query. Throughout both stages, we freeze the vision encoder and visual feature mergers and apply LoRA~\citep{lora} only to the text decoder. Following the merge-and-reinitialize principle of ReLoRA~\citep{relora}, we merge the first-stage LoRA updates into the backbone and initialize fresh adapters for the second stage. Detailed data preparation procedures and training hyperparameters are provided in the Appendix~\ref{sec:training_details}.

\paragraph{Evaluation}
We evaluate RenderRank on 11 BEIR datasets~\citep{beir}: ArguAna~\citep{AA}, Climate-FEVER~\citep{CFV}, DBPedia~\citep{DBP}, FiQA~\citep{FQA}, FEVER~\citep{FVR}, HotpotQA~\citep{HQA}, NFCorpus~\citep{NFC}, SCIDOCS~\citep{SD}, SciFact~\citep{SF}, TREC-COVID~\citep{TC}, and Touché-2020~\citep{TCH}. For each BEIR query, we rerank the top 100 documents retrieved by BM25. To assess long-document reranking performance, we evaluate on the English subset of MLDR~\citep{bge} following the MMTEB~\citep{mmteb}, and on three LongEmbed datasets~\citep{zhu2024longembed}: 2WikiMQA, QMSum, and SummScreenFD. For the latter, we retrieve the top eight documents per query using Qwen3-Embedding-0.6B~\citep{qwen3embedding} and rerank them, consistent with the eight-candidate evaluation protocol used for MLDR. We use nDCG@10 as the evaluation metric. All baseline rerankers use text queries and documents. For RenderRank, queries remain in text form, while documents are rendered as images following Section~\ref{sec:RenderRank}. Consistent with prior work on visual feature precomputation~\citep{taraday2025efficient,fan2026minireranker}, We assume that visual embeddings of document images have been computed in advance.

\paragraph{Baselines}
We compare RenderRank with rerankers spanning diverse architectures and model sizes: gte-reranker-modernbert-base~\citep{gte}, mxbai-rerank-large-v1 and v2~\citep{mxbaiv1,mxbaiv2}, bge-reranker-large and bge-reranker-v2-gemma~\citep{bge_reranker,bge}, Qwen3-Reranker-0.6B and 4B~\citep{qwen3embedding}, LAMAR-600m~\citep{lamar}, llama-nemotron-rerank-1b-v2, LightOn-rerank-PW-2B and 4B~\citep{lighton}, and zerank-2-reranker~\citep{zerank}.

%% file: Texs/Results.tex
\section{Experimental Results}
\subsection{Reranking Performance}
\input{Tables/Main}
Table~\ref{tab:beir_main} compares the reranking performance and average input token counts of RenderRank and a range of text-based rerankers across 11 BEIR benchmarks. RenderRank achieves an average NDCG@10 of 55.96, demonstrating competitive performance against text-based rerankers. For example, it outperforms zerank-2-reranker and LightOn-rerank-PW-4B by 7.66\% and 7.26\% in average NDCG@10, respectively. This performance is obtained with an average of 290.07 input tokens per query--document pair, 16.5--35.5\% fewer than the text-based rerankers evaluated. These results show that representing document text as visual tokens reduces the number of tokens processed by the language backbone while effectively scoring document relevance to the query. Such token savings are particularly relevant to reranking, where each candidate document is scored separately with the query, so the reduction in input tokens applies to every candidate evaluation.

\subsection{Efficiency Evaluation}
\label{sec:ee}
To evaluate the extent to which fewer input tokens reduce computational cost and improve reranking throughput, we analyze both estimated computation and measured inference throughput. Shorter inputs can reduce the computational cost of scoring each candidate, but this does not necessarily lead to proportional gains in throughput. We therefore examine these two aspects separately. Specifically, we (1) compare computational costs based on model architecture and input length, and (2) measure the number of pairs processed per second to assess RenderRank's inference efficiency.

\input{Tables/Flops}
\subsubsection{Computational Efficiency}
\paragraph{Setup.}
We estimate computational cost using the approximation in E$^2$R-FLOPs~\citep{peng-etal-2025-efficiency}, accounting for model architecture and individual input lengths. The computational cost of a single forward pass through a standard dense Transformer is approximated as follows:
\begin{equation}
    \begin{gathered}
    F(n)
    =
    4Lh\left[\left(1+\frac{1}{r}\right)h+f\right]n
    +
    \frac{4Lhn^2}{r},
    \\[-2pt]
    \scriptstyle
    L:\text{ layers},\quad
    h:\text{ hidden dim},\quad
    f:\text{ FFN dim},\quad
    n:\text{ input length},\quad
    r=H_Q/H_{KV}:\text{ Q/KV head ratio}
    \end{gathered}
\end{equation}
We adjust this expression according to each model's architecture and attention pattern. Let $n_{q,d}$ denote the actual number of input tokens for query $q$ and document $d$ after truncation to the maximum sequence length, and compute the corresponding cost as $C_{q,d}=F(n_{q,d})$. We define $C_q$ as the sum of these costs across all candidates for a query. Here, $\mathrm{AVG}_{C_{q,d}}$ denotes the average cost across all evaluated pairs, whereas $\mathrm{AVG}_{C_q}$ denotes the average total cost per query. TFLOPs and QPP (Queries per PetaFLOP) are calculated as follows:
\begin{equation}
\mathrm{TFLOPs/pair}
=
\frac{\mathrm{AVG}_{C_{q,d}}}{10^{12}}
\qquad
\mathrm{QPP}
=
\frac{1}{\mathrm{AVG}_{C_q}/10^{15}}
\end{equation}
TFLOPs measures the average computation required to score the relevance of an individual pair, with lower values indicating lower per-pair computational cost. QPP measures the number of queries for which all candidate documents can be reranked within a fixed computational budget, representing processing capacity per unit of computation rather than processing speed over time.

\paragraph{Results.}
Table~\ref{tab:model_architecture} compares model architectures and estimated computational costs. RenderRank requires an average of 0.63 TFLOPs per query--document pair and achieves an average QPP of 20.1. For example, compared with bge-reranker-v2-gemma, which requires 1.10 TFLOPs per pair and achieves a QPP of 12.3, RenderRank offers comparable reranking performance at a lower computational cost. Its higher QPP indicates that more queries can have their candidate sets reranked within the same computational budget. For a given model, shorter sequences reduce the computational cost of attention as well as the projections and FFNs applied to each token. Document compression through visual tokens can therefore lower the computational cost per candidate even in reranking, where relevance is scored in a single forward pass. However, differences in computational cost across models also reflect backbone size and attention architecture, which should be considered when assessing the contribution of reduced input length.

\begin{wrapfigure}{r}{0.42\columnwidth}
    \vspace{-15pt}
    \centering
    \includegraphics[width=\linewidth]{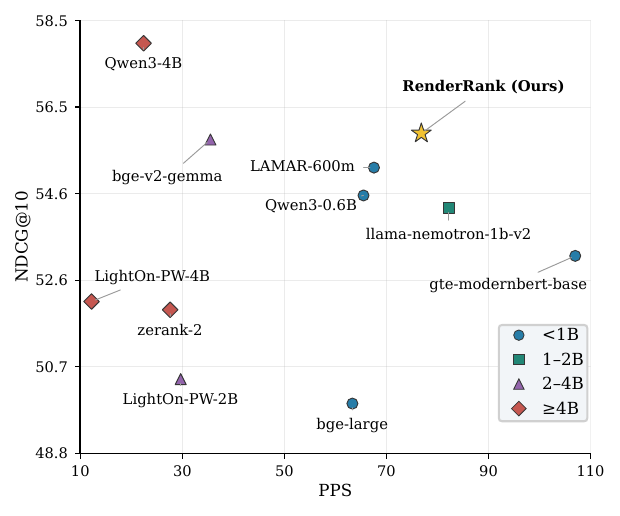}
    \caption{Reranking effectiveness and GPU forward throughput, averaged over 11 BEIR benchmarks. For each model and benchmark, PPS is measured at the batch size yielding the highest throughput.}
    \label{fig:efficiency}
    \vspace{-30pt}
\end{wrapfigure}

\subsubsection{Inference Throughput}
\paragraph{Setup.}
To evaluate reranking throughput in practice, we measure Pairs per Second (PPS) following the measurement procedure of~\citep{aarsen2026ettin-reranker}. Experiments are conducted on an NVIDIA A6000 48GB using the maximum sequence length supported by each model. Starting at 8, we double the batch size until an out-of-memory (OOM) error occurs or the search limit is reached, testing smaller batch sizes when necessary. Let $\mathcal{B}$ denote the set of successfully measured batch sizes, $N$ the total number of evaluated pairs, and $T_b$ the total GPU forward time at batch size $b$. We report the highest PPS across successfully measured batch sizes: 
\begin{equation}
\mathrm{PPS}
=
\max_{b\in\mathcal{B}}\frac{N}{T_b}
\end{equation}
$T_b$ is measured in seconds using CUDA events after warm-up at each batch size and includes only GPU forward time, excluding preprocessing and cache I/O.

\paragraph{Results.}
Figure~\ref{fig:efficiency} compares average NDCG@10 and PPS across models. RenderRank achieves an average throughput of 76.83 PPS, exceeding that of larger models and several similarly sized models. For example, it achieves approximately twice the throughput of bge-reranker-v2-gemma at comparable reranking performance. The throughput advantage also extends to some smaller models. Despite its higher estimated TFLOPs per pair, RenderRank exceeds the throughput of 67.55 PPS measured for LAMAR-600m. This result indicates that lower estimated TFLOPs per pair do not necessarily correspond to higher inference throughput. RenderRank supports higher candidate evaluation throughput while maintaining high ranking performance.

\input{Tables/longdoc}

\subsection{Long Document Reranking}

Table~\ref{tab:long_document_results} compares reranking performance, average input token counts, and throughput (PPS) on four long-document datasets. RenderRank achieves an average NDCG@10 of 88.27, demonstrating strong reranking performance in prioritizing relevant documents even for long inputs. On MLDR, it achieves an NDCG@10 of 99.74 and uses an average of 4.20K input tokens, approximately half that of the text-based rerankers evaluated. This reduction extends to the three LongEmbed datasets, where RenderRank uses 53--57\% fewer input tokens than the most token-efficient text baseline. RenderRank achieves the highest throughput among the compared models on all four datasets, averaging 4.51 PPS versus 2.66 PPS for the fastest baseline, a $1.70\times$ improvement. The corresponding speedups range from $1.55\times$ to $1.91\times$ across individual datasets. For long inputs, attention computation grows with sequence length, alongside the cost of token-wise projections and FFNs; shorter input representations can therefore reduce the computational cost even when scoring an individual pair. These results show that RenderRank provides high ranking quality for long documents while reducing input token counts and improving inference speed.

%% file: Tables/Main.tex
\begin{table*}[t]
\centering
\caption{Reranking performance (NDCG@10) and input token counts on 11 BEIR datasets. Avg.\ and Tokens report the mean performance and input length per query--document pair, respectively, across datasets. Params counts all model parameters. For RenderRank, input length includes both textual and visual tokens processed by the language backbone. Bold and underlined values indicate the top two results across rerankers, respectively.}

\label{tab:beir_main}
\scriptsize
\renewcommand{\arraystretch}{1.05}
\setlength{\tabcolsep}{3.2pt}
\resizebox{\textwidth}{!}{
\begin{tabular}{lcccccccccccccc}
\toprule
& & \multicolumn{12}{c}{\textbf{BEIR}} 
& \multirow{2}{*}{\textbf{Tokens} $\downarrow$} \\
\cmidrule(lr){3-14}
\textbf{Model}
& \textbf{Params}
& \textbf{AA}
& \textbf{CFV}
& \textbf{DBP}
& \textbf{FQA}
& \textbf{FVR}
& \textbf{HQA}
& \textbf{NFC}
& \textbf{SD}
& \textbf{SF}
& \textbf{TC}
& \textbf{TCH}
& \textbf{Avg.}
& \\
\midrule

BM25
& --
& 39.70 & 16.51 & 31.83 & 23.61 & 65.18 & 63.30
& 32.18 & 14.90 & 67.89 & 59.47 & 44.22
& 41.71 & -- \\

\midrule
\multicolumn{15}{l}{\textit{Rerankers with $<2$B parameters}} \\
\midrule

gte-reranker-modernbert-base
& 150M
& 66.14 & 25.80 & 42.10 & 42.54 & \textbf{89.84} & 76.15
& 34.81 & 18.90 & 75.81 & 79.66 & 33.44
& 53.20 & 359.25 \\

mxbai-rerank-large-v1
& 435M
& 16.16 & 25.61 & 45.37 & 40.29 & 82.19 & 71.58
& 37.32 & 18.96 & 75.26 & \textbf{85.84} & 37.44
& 48.73 & \underline{347.45} \\

bge-reranker-large
& 560M
& 30.94 & 34.11 & 44.21 & 37.47 & 89.30 & 80.09
& 33.91 & 16.68 & 73.98 & 73.21 & 34.69
& 49.87 & 409.32 \\

LAMAR-600m
& 560M
& 66.21 & \underline{36.21} & \underline{46.23} & 41.14 & \underline{89.49} & 79.14
& 35.04 & 20.22 & 76.97 & 80.65 & 35.76
& 55.19 & 409.32 \\

Qwen3-Reranker-0.6B
& 600M
& 68.04 & 34.34 & 43.76 & 39.74 & 87.13 & 77.25
& 36.60 & 20.44 & 77.00 & \underline{85.54} & 30.32
& 54.56 & 438.70 \\

llama-nemotron-rerank-1b-v2
& 1.2B
& 54.29 & 27.37 & 45.13 & \textbf{47.11} & 87.66 & \underline{80.34}
& \underline{38.14} & \underline{21.53} & \textbf{79.82} & 83.14 & 32.47
& 54.27 & 363.99 \\

mxbai-rerank-large-v2
& 1.5B
& 42.23 & 23.03 & 38.01 & 28.85 & 70.84 & 69.50
& 35.17 & 17.49 & 78.81 & 67.78 & \textbf{46.94}
& 47.15 & 449.76 \\

\midrule
\multicolumn{15}{l}{\textit{Rerankers with $\geq 2$B parameters}} \\
\midrule


LightOn-rerank-PW-2B
& 2.2B
& 45.52 & 23.21 & 42.60 & 35.66 & 86.51 & 73.05
& 35.61 & 16.64 & 77.70 & 81.24 & 36.91
& 50.42 & 416.22 \\

bge-reranker-v2-gemma
& 2.5B
& \underline{74.98} & 32.03 & 45.01 & 43.18 & 87.45 & \textbf{80.47}
& 36.25 & 19.58 & 77.43 & 80.08 & 37.56
& 55.82 & 399.11 \\

Qwen3-Reranker-4B
& 4B
& \textbf{75.40} & \textbf{37.84} & \textbf{47.02} & 44.63 & 89.14 & 79.06
& 37.48 & \textbf{23.59} & \underline{79.47} & 85.47 & \underline{38.77}
& \textbf{57.99} & 438.70 \\

zerank-2-reranker
& 4B
& 44.87 & 23.65 & 44.95 & 42.75 & 83.22 & 70.97
& \textbf{38.50} & 20.23 & 79.35 & 85.24 & 38.10
& 51.98 & 380.69 \\

LightOn-rerank-PW-4B
& 4.5B
& 52.50 & 28.24 & 44.80 & 40.27 & 87.70 & 72.79
& 37.33 & 17.29 & 76.72 & 79.54 & 36.70
& 52.17 & 416.22 \\

\midrule

\textbf{RenderRank (Ours)}
& 2.1B
& 69.11 & 35.91 & 46.19 & 42.66 & 88.92 & 78.60
& 37.31 & 20.75 & 77.00 & 84.82 & 34.27
& \underline{55.96}
& \textbf{290.07} \\

\bottomrule
\end{tabular}
}

\end{table*}

%% file: Tables/Flops.tex
\begin{table*}[t]
\centering
\caption{Reranker architectures and estimated computational efficiency, with input token counts and efficiency metrics averaged over 11 BEIR datasets. $Q$, $KV$, and $d_h$ denote the numbers of query and key--value heads and the head dimension, respectively. $G$, $L$, $SW$, and $Lin$ denote global, local, sliding-window, and linear attention. For RenderRank, Enc.\ and Dec.\ denote the vision encoder and language decoder. Its parameter count includes all model components, whereas TFLOPs and QPP exclude offline vision encoding. Bold and underlined values indicate the top two results within each parameter group, respectively. $^\ast$GQA applies only to the global-attention layers.}

\label{tab:model_architecture}

\scriptsize
\setlength{\tabcolsep}{2.8pt}
\renewcommand{\arraystretch}{1.0}
\resizebox{\textwidth}{!}{
\begin{tabular}{
l c c c c c c c c c c 
>{\columncolor{gray!15}}c
>{\columncolor{gray!15}}c
>{\columncolor{gray!15}}c
}
\toprule

& \multicolumn{10}{c}{\textbf{Architecture}}
& \multicolumn{3}{c}{\textbf{Efficiency}} \\
\cmidrule(lr){2-11}\cmidrule(lr){12-14}

\textbf{Model}
& \textbf{Params}
& \textbf{Layers}
& \textbf{Hidden}
& \textbf{FFN}
& \textbf{Q}
& \textbf{KV}
& $\boldsymbol{d_h}$
& \textbf{Bi.}
& \textbf{Attn.}
& \textbf{Attn. Pattern}
& \multicolumn{1}{c}{\textbf{Tokens} $\downarrow$}
& \multicolumn{1}{c}{\textbf{TFLOPs} $\downarrow$}
& \multicolumn{1}{c}{\textbf{QPP} $\uparrow$} \\

\midrule
\multicolumn{14}{l}{\textit{Rerankers with $<2$B parameters}} \\
\midrule

\rule{0pt}{2.2ex}gte-reranker-modernbert-base
& 150M
& 22
& 768
& 1152
& 12
& 12
& 64
& O
& MHA
& G8+L14
& \underline{359.25}
& \textbf{0.07}
& \textbf{209.5} \\

mxbai-rerank-large-v1
& 435M
& 24
& 1024
& 4096
& 16
& 16
& 64
& O
& MHA
& G24
& \textbf{347.45}
& \underline{0.18}
& \underline{73.4} \\

bge-reranker-large
& 560M
& 24
& 1024
& 4096
& 16
& 16
& 64
& O
& MHA
& G24
& 409.32
& 0.21
& 63.3 \\

LAMAR-600m
& 560M
& 24
& 1024
& 4096
& 16
& 16
& 64
& O
& MHA
& G24
& 409.32
& 0.28
& 56.8 \\

Qwen3-Reranker-0.6B
& 600M
& 28
& 1024
& 3072
& 16
& 8
& 128
& X
& GQA
& G28
& 438.70
& 0.25
& 53.2 \\

llama-nemotron-rerank-1b-v2
& 1.2B
& 16
& 2048
& 8192
& 32
& 8
& 64
& O
& GQA
& G16
& 363.99
& 0.52
& 28.2 \\

mxbai-rerank-large-v2
& 1.5B
& 28
& 1536
& 8960
& 12
& 2
& 128
& X
& GQA
& G28
& 449.76
& 0.84
& 15.2 \\

\midrule
\multicolumn{14}{l}{\textit{Rerankers with $\geq 2$B parameters}} \\
\midrule



LightOn-rerank-PW-2B
& 2.2B
& 24
& 2048
& 6144
& 8
& 2
& 256
& X
& GQA$^\ast$
& G6+Lin18
& 416.22
& \underline{0.73}
& \underline{18.3} \\

bge-reranker-v2-gemma
& 2.5B
& 18
& 2048
& 16384
& 8
& 1
& 256
& X
& MQA
& G18
& 399.11
& 1.10
& 12.3 \\

Qwen3-Reranker-4B
& 4B
& 36
& 2560
& 9728
& 32
& 8
& 128
& X
& GQA
& G36
& 438.70
& 2.12
& 6.1 \\

zerank-2-reranker
& 4B
& 36
& 2560
& 9728
& 32
& 8
& 128
& X
& GQA
& G36
& \underline{380.69}
& 1.84
& 7.8 \\

LightOn-rerank-PW-4B
& 4.5B
& 32
& 2560
& 9216
& 16
& 4
& 256
& X
& GQA$^\ast$
& G8+Lin24
& 416.22
& 1.73
& 7.7 \\

\specialrule{\lightrulewidth}{0pt}{0pt}

\multirow{2}{*}{\textbf{RenderRank (Ours)}}
& \multirow{2}{*}{2.1B}
& \rule{0pt}{2.5ex}24 (Enc.)
& 1024
& 4096
& 16
& 16
& 64
& O
& MHA
& G24
& & & \\

&
& 28 (Dec.)
& 2048
& 6144
& 16
& 8
& 128
& X
& GQA
& G28
& \multirow{-2}{*}{\textbf{290.07}}
& \multirow{-2}{*}{\textbf{0.63}}
& \multirow{-2}{*}{\textbf{20.1}} \\

\bottomrule
\end{tabular}
}

\end{table*}

%% file: Tables/longdoc.tex
\begin{table*}[t]
    \centering
    \caption{Reranking performance, average input token counts, and throughput on four long-document datasets. The comparison includes models officially supporting sequence lengths $\geq 16$K tokens.}
    \label{tab:long_document_results}
    
    \scriptsize
    \renewcommand{\arraystretch}{1.25}
    \setlength{\tabcolsep}{3pt}
    \setlength{\aboverulesep}{0pt}
    \setlength{\belowrulesep}{0pt}
    
    \resizebox{\textwidth}{!}{
    \begin{tabular}{l*{12}{c}|cc}
    \toprule
    \multirow{3}{*}{\textbf{Model}}
    & \multicolumn{3}{c}{\multirow{2}{*}{\textbf{MLDR}}}
    & \multicolumn{9}{c}{\textbf{LongEmbed}}
    & \multicolumn{2}{c}{\multirow{2}{*}{\textbf{Avg.}}} \\
    \cmidrule(lr){5-13}
    & \multicolumn{3}{c}{}
    & \multicolumn{3}{c}{\textbf{2WikiMQA}}
    & \multicolumn{3}{c}{\textbf{QMSum}}
    & \multicolumn{3}{c}{\textbf{SummScreenFD}}
    & \multicolumn{2}{c}{} \\
    \cmidrule(lr){2-4}
    \cmidrule(lr){5-7}
    \cmidrule(lr){8-10}
    \cmidrule(lr){11-13}
    \cmidrule(lr){14-15}
    
    & \textbf{N@10} & \textbf{Tokens} & \textbf{PPS}
    & \textbf{N@10} & \textbf{Tokens} & \textbf{PPS}
    & \textbf{N@10} & \textbf{Tokens} & \textbf{PPS}
    & \textbf{N@10} & \textbf{Tokens} & \textbf{PPS}
    & \textbf{N@10} & \textbf{PPS} \\
    \midrule
    
    Qwen3-Reranker-0.6B
    & 99.63 & 8863.4 & \underline{2.59}
    & \cellcolor{gray!15}\textbf{94.54} & \cellcolor{gray!15}9230.2 & \cellcolor{gray!15}\underline{2.23}
    & 56.43 & 13543.6 & \underline{1.60}
    & \cellcolor{gray!15}98.25 & \cellcolor{gray!15}8606.3 & \cellcolor{gray!15}\underline{4.21}
    & 87.21 & \underline{2.66} \\
    
    LightOn-rerank-PW-2B
    & 98.91 & 8878.0 & 1.10
    & \cellcolor{gray!15}71.27 & \cellcolor{gray!15}9230.1 & \cellcolor{gray!15}0.90
    & 54.30 & 13876.6 & 0.57
    & \cellcolor{gray!15}93.05 & \cellcolor{gray!15}8977.2 & \cellcolor{gray!15}1.21
    & 79.38 & 0.94 \\
    
    Qwen3-Reranker-4B
    & \textbf{99.85} & 8863.4 & 0.87
    & \cellcolor{gray!15}\textbf{94.54} & \cellcolor{gray!15}9230.2 & \cellcolor{gray!15}0.76
    & 59.11 & 13543.6 & 0.45
    & \cellcolor{gray!15}\textbf{99.11} & \cellcolor{gray!15}8606.3 & \cellcolor{gray!15}1.13
    & 88.15 & 0.80 \\
    
    zerank-2-reranker
    & 99.57 & \underline{8805.6} & 0.88
    & \cellcolor{gray!15}\underline{94.42} & \cellcolor{gray!15}\underline{9173.1} & \cellcolor{gray!15}0.61
    & \textbf{60.10} & \underline{13485.9} & 0.34
    & \cellcolor{gray!15}98.97 & \cellcolor{gray!15}\underline{8548.5} & \cellcolor{gray!15}1.12
    & \underline{88.26} & 0.74 \\
    
    LightOn-rerank-PW-4B
    & 99.68 & 8878.0 & 0.53
    & \cellcolor{gray!15}93.84 & \cellcolor{gray!15}9230.1 & \cellcolor{gray!15}0.44
    & 58.87 & 13876.6 & 0.26
    & \cellcolor{gray!15}98.91 & \cellcolor{gray!15}8977.2 & \cellcolor{gray!15}0.52
    & 87.82 & 0.44 \\
    
    \midrule
    
    \textbf{RenderRank (Ours)}
    & \underline{99.74} & \textbf{4198.0} & \textbf{4.59}
    & \cellcolor{gray!15}94.30 & \cellcolor{gray!15}\textbf{4262.2} & \cellcolor{gray!15}\textbf{4.25}
    & \underline{59.94} & \textbf{6388.8} & \textbf{2.67}
    & \cellcolor{gray!15}\underline{99.10} & \cellcolor{gray!15}\textbf{3701.9} & \cellcolor{gray!15}\textbf{6.52}
    & \textbf{88.27} & \textbf{4.51} \\
    
    \bottomrule
    \end{tabular}
    }
    \end{table*}

%% file: Texs/Analysis_Ablation.tex
\section{Analysis and Ablation Studies}

\input{Tables/Budget}
\subsection{Reranking under Fixed Token Budgets}
To evaluate whether visual document representations can preserve more document content within the same length limit and improve reranking performance, we evaluate all models on the same MLDR samples, setting the maximum sequence length to 2K, 4K, and 8K in separate evaluations. Table~\ref{tab:mldr_length} shows that RenderRank achieves the highest NDCG@10 across all three settings. At 2K, RenderRank achieves an NDCG@10 of 97.9, surpassing the scores of 95.8 and 97.3 obtained by gte-reranker and LightOn-PW-4B at 4K, respectively. At 4K, it reaches 99.5, comparable to the performance of text-based rerankers at 8K, providing strong long-document reranking performance with half the maximum sequence length. Performance differences between models narrow at 8K, whereas differences across document representations are more pronounced at shorter lengths. These results suggest that RenderRank can encode and leverage more document content within the same length limit for long-document reranking, providing ranking quality comparable to text-based rerankers with fewer input tokens.

\begin{wrapfigure}{r}{0.38\columnwidth}
    \vspace{-13pt}
    \centering
    \includegraphics[width=\linewidth]{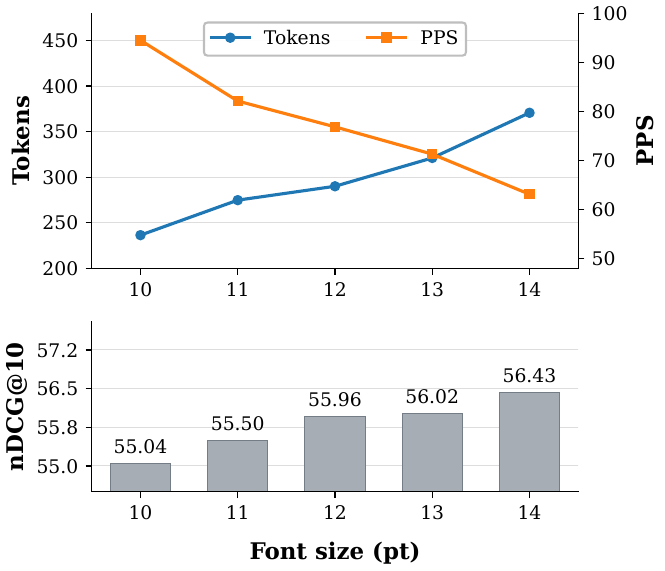}
    \caption{Effect of rendering font size on reranking performance, input token counts, and throughput, averaged over 11 BEIR datasets.}
    \label{fig:font_size}
    \vspace{-14pt}
\end{wrapfigure}

\subsection{Effect of Rendering Font Size}
We vary the rendering font size to examine how the visual information density of documents affects reranking quality and inference efficiency. As shown in Figure~\ref{fig:font_size}, increasing the font size from 10pt to 14pt raises the average input length from approximately 236 to 370 tokens and reduces throughput from approximately 94 to 63 PPS, while improving NDCG@10 from 55.04 to 56.43. Smaller fonts fit more content within the same image area, shortening the input sequence, but allocate fewer pixels to each character, potentially making details needed for relevance assessment harder to distinguish. The default 12pt setting achieves a score of 55.96 with approximately 290 input tokens and 77 PPS, offering fewer tokens and higher throughput than 14pt with a relative difference of approximately 0.83\%. These results show that document compression through visual tokens can improve throughput while allowing the balance between ranking quality and inference efficiency to be adjusted through the degree of compression.

\input{Tables/Ablation}
\subsection{Component Ablation}

We examine how the training stages and inference settings affect reranking performance and throughput, respectively. As shown in Table~\ref{tab:ablation}(a), Cross-Modal Relevance Distillation improves average NDCG@10 with image inputs from 50.20 to 54.86. This improvement demonstrates the contribution of learning relevance scores from a text-based teacher to reranking with visual inputs. Query-Local Relevance Discrimination further increases performance to 55.96, showing the additional benefit of learning relative priorities among candidates after distillation. Table~\ref{tab:ablation}(b) compares inference throughput across input and caching settings. Cached image inputs achieve 76.83 PPS, compared with 37.31 PPS when visual encoding is included in the measurement. In settings where document visual embeddings are stored in advance, reranking can directly use these representations to improve throughput. These results show that the two training stages improve relevance scoring for documents represented as compressed visual tokens, while inference throughput varies with the execution setting even for the same model.

%% file: Tables/Budget.tex
\begin{wraptable}{r}{0.39\columnwidth}
    \vspace{-12pt}
    \centering
    \caption{Reranking performance on MLDR under different maximum sequence lengths. Bold indicates the best result at each length.}
    \label{tab:mldr_length}
    \small
    \renewcommand{\arraystretch}{1.0}
    \setlength{\tabcolsep}{2.7pt}
    \resizebox{\linewidth}{!}{%
    \begin{tabular}{lccc}
        \toprule
        Model & 2K & 4K & 8K \\
        \midrule
          gte-reranker-modernbert & 93.2 & 95.8 & 98.5 \\
          llama-nemotron-rerank-1b-v2  & 97.4 & 98.5 & 99.4 \\
          Qwen3-Reranker-4B           & 96.9 & 98.0 & 99.6 \\
          LightOn-rerank-PW-4B        & 95.1 & 97.3 & 99.5 \\
          \midrule
          RenderRank & \textbf{97.9} & \textbf{99.5} & \textbf{99.7} \\
      \bottomrule
    \end{tabular}%
    }
    \vspace{-12pt}
\end{wraptable}

%% file: Tables/Ablation.tex
\begin{wraptable}{r}{0.32\columnwidth}
\vspace{-13pt}
\centering
\caption{Training-stage effectiveness and inference throughput on 11 BEIR datasets. Effectiveness is averaged across datasets, and throughput is measured using a 12pt font.}
\label{tab:ablation}
\scriptsize
\renewcommand{\arraystretch}{1.}
\resizebox{\linewidth}{!}{%
\begin{tabular}{l@{\hspace{30pt}}c}
    \toprule
    \multicolumn{2}{c}{\emph{(a) Reranking effectiveness}} \\[-2pt]
    \cmidrule{1-2}

    \textbf{Model}
    & \textbf{N@10} \\
    RenderRank
    & \textbf{55.96} \\
    \cdashline{1-2}[0.5pt/0.8pt]
    \quad w/o Stage 2
    & 54.86 \\
    \quad w/o Stage 1 \& 2
    & 50.20 \\
    \addlinespace[2pt]
    \specialrule{0.8pt}{0pt}{2pt}
    \multicolumn{2}{c}{\emph{(b) Inference throughput}} \\[-2pt]
    \cmidrule{1-2}

    \textbf{Inference setting}
    & \textbf{PPS} $\uparrow$ \\
    RenderRank
    & \textbf{76.83} \\
    \cdashline{1-2}[0.5pt/0.8pt]
    \quad w/o caching
    & 37.31 \\
    \bottomrule
\end{tabular}%
}
\end{wraptable}

%% file: Texs/Conclusion.tex
\section{Conclusion}
We presented RenderRank, which renders document text as images and encodes it into compressed visual tokens for query-dependent relevance scoring. Cross-modal distillation and query-local discrimination enable the model to learn relevance from compressed visual inputs and refine the relative scores of candidate documents. RenderRank achieves high reranking performance with fewer input tokens, while its compressed representations allow more document content to contribute to relevance scoring within the same sequence length limit. Our analyses show that both training stages improve relevance scoring, while rendering density controls the trade-off between reranking performance and inference efficiency. This work establishes compressed visual representations as an effective basis for document reranking and opens a path toward representing and reranking longer text documents.

%% file: Texs/Appendix.tex
\newpage

\section{Further Related Work}
\label{app:further_related_work}

Learned compressed representations have been explored as substitutes for document token sequences to reduce the computational cost of reranking~\citep{dejean2026efficient}. In this approach, a separate compressor encodes each document into representations associated with a fixed number of memory tokens, and the compressor and reranker are jointly trained with a ranking objective. After training, document representations are computed in advance, allowing the reranker to process the compressed representations alongside the query instead of the original text.

Other approaches reduce token computation within the reranker. One approach encodes documents independently, selects the document key--value states accessible to query-side attention, and directly supervises relevance scores derived by aggregating attention weights~\citep{lu2026comprank}. Another pools token representations at an intermediate layer to shorten the sequences processed by subsequent layers, training the reranker to adapt to the compressed representations~\citep{zhuang2026layer}. Subsequent work trains a single model across multiple compression ratios, allowing the ratio to be adjusted at inference time~\citep{wang2026tevatron}.

Rather than jointly training a text compressor with the reranker or selecting or merging tokens within the language model used for reranking, RenderRank represents rendered documents through the visual encoder of an existing VLM. We freeze the visual encoder and feature mergers and train the language model for relevance scoring, with the number of visual tokens varying according to the rendering configuration and document length. Our focus is on adapting existing visual text representations to document reranking through query-dependent relevance training, and on characterizing the resulting trade-offs between reranking performance and inference efficiency.

\input{Aappendix_Tables/App_gen}

\section{Generation Experiments for Rendering Evaluation}
\label{app:rendering_evaluation}
We evaluate rendering configurations in terms of task performance and token efficiency using Gemma-4-26B-A4B-it~\citep{gemmateam2026gemma4} and Qwen3.6-35B-A3B~\citep{qwen36_35b_a3b} on the QASPER~\citep{qasper}, GovReport~\citep{govreport}, and Multi-News~\citep{multinews} datasets in LongBench~\citep{bai2024longbench}. We report F1 for QASPER and ROUGE-L for both summarization tasks. Documents are rendered in Roboto Regular, combining font sizes of 10pt, 11pt, and 12pt with line spacing of 1.0, 1.1, and 1.2. Images have a fixed width of 896 pixels and a height that adjusts to the content, up to 896 pixels. Content exceeding this height continues onto the next image without overlap, and the resulting images are supplied in document order. We use the recommended sampling settings for each model and keep them unchanged across all rendering configurations and the text baseline. Thinking mode is disabled for both models.

Table~\ref{tab:rendering_config_full} supplements the main-text results with average document-token counts and an additional evaluation on Multi-News. Document-token counts exclude queries and instructions, and token reduction is calculated relative to the text baseline for each model and dataset. The additional results show that changes in token count do not consistently translate into corresponding changes in performance. For example, increasing line spacing from 1.0 to 1.2 at 12pt improves the Multi-News score of Qwen from 22.98 to 23.41, but reduces token savings from 42.66\% to 27.92\%. In contrast, Gemma at 12pt with line spacing of 1.2 uses more document tokens than the text baseline without reaching its performance. These results highlight that the effects of rendering configurations vary across models and tasks, underscoring the importance of considering both performance and token efficiency when determining the rendering configuration.

\begin{table}[t]
\centering
\scriptsize
\caption{An example from Touché-2020 showing the query, document text, and two rendered images.}
\label{tab:rendering_example}
\renewcommand{\arraystretch}{1.2}
\begin{tabular}{@{}p{\linewidth}@{}}
\toprule
\textbf{Query:} Should election day be a national holiday? \\
\midrule
\textbf{Document:} Autumn Regular Tournament: In a democracy, voting ought to be compulsory Thanks, Philo. =Pro case=FrameworkPro argues that the only concern that should govern policy is if that policy is "democratic". This is absurd--democracies are governments and as I articulated in my framework, governance is fundamentally a balancing act between individual rights and the common good. Good governments are ones which make acceptable trade offs on this scale. ... The only truly neutral act is to not vote. Pro does not dispute that we should use the Sherbert Test: forcing people to violate their religion and hoping that they'll know to spoil their ballots is not the *least intrusive* means of achieving high turnout. Allowing them to abstain and encouraging voluntary voting through tax credits is.Pro removes legitimitate means of political expression and forces us all to give consent to the system. I'll discuss the rest of my case and crystallize the debate in the next round. 1. http://tinyurl.com...2. http://tinyurl.com...3. http://tinyurl.com... \\
\midrule
\begin{minipage}[t]{0.5\linewidth}
  \centering
  \textbf{Image 1}\par\smallskip
  \includegraphics[width=\linewidth]{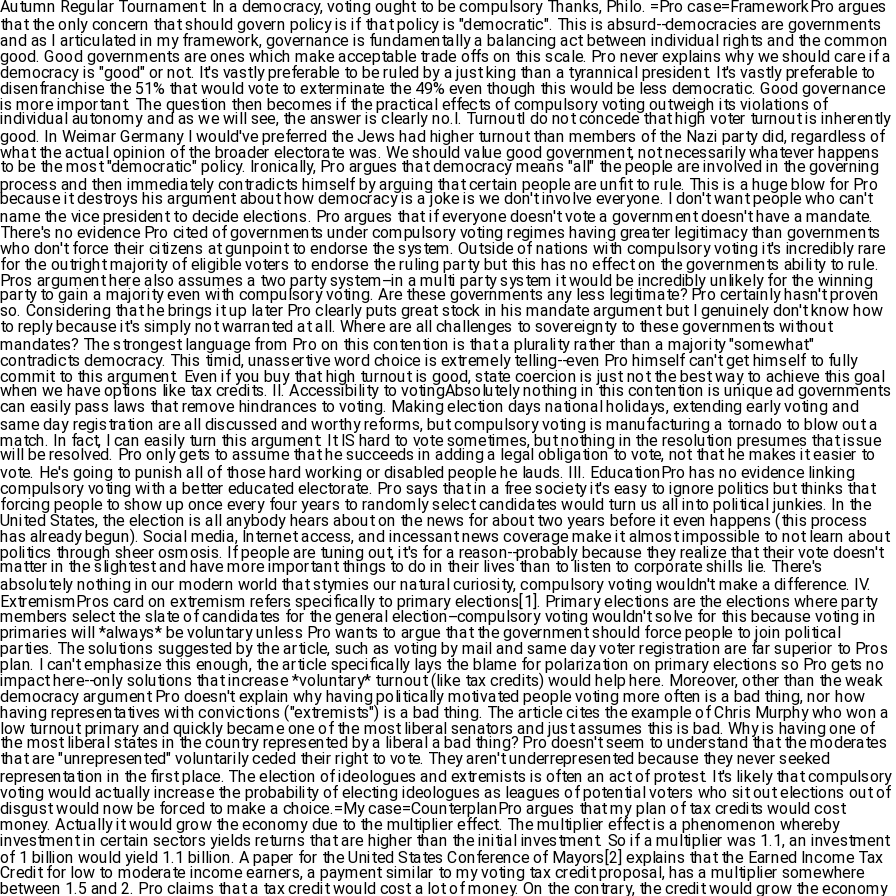}
\end{minipage}%
\begin{minipage}[t]{0.5\linewidth}
  \centering
  \textbf{Image 2}\par\smallskip
  \includegraphics[width=\linewidth]{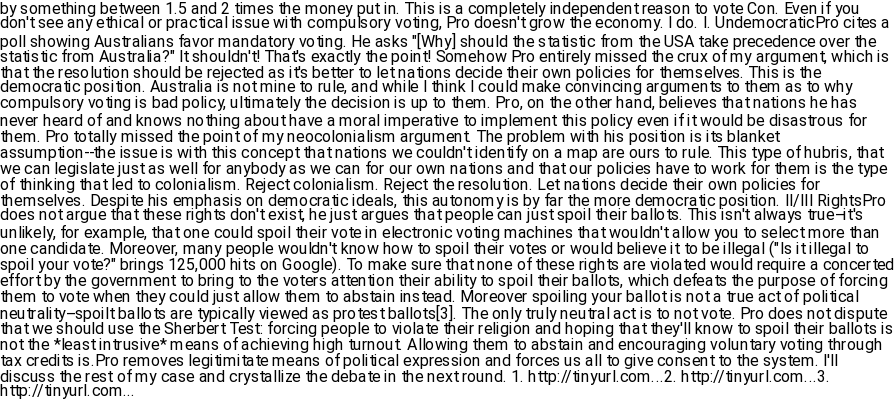}
\end{minipage} \\
\bottomrule
\end{tabular}
\end{table}

\section{Training Details}
\label{sec:training_details}
\paragraph{Rendering}
We render document text into images while retaining each query as text. Before rendering, all whitespace characters, including line breaks, tabs, and consecutive spaces, are normalized to a single space; original paragraph boundaries and manual line breaks are therefore not preserved. Documents are rendered in black text on a white background using Roboto Regular at 12\,pt (16\,px at 96\,DPI), with zero margins and Pillow's default text antialiasing. Each image has a fixed width of 896\,px. Text is wrapped according to its measured pixel width under the selected font, rather than a fixed character count, and words exceeding the image width are split at character boundaries.

We use a line spacing of 1.0, corresponding to a 16\,px vertical advance between consecutive lines. Each image contains at most 56 lines, yielding a maximum height of 896\,px. Longer documents continue onto subsequent images without overlap or repeated text. The height of each image is rounded up to the nearest multiple of 32\,px, with a minimum of 32\,px and a maximum of 896\,px; thus, the final image is cropped in height to accommodate its remaining lines rather than padded to a full square. Table~\ref{tab:rendering_example} presents an example query--document pair and its corresponding rendered document images.

\input{Aappendix_Tables/dataset}

\paragraph{Dataset}
Table~\ref{tab:training_data} summarizes the source dataset composition for both training stages. For Cross-Modal Relevance Distillation in Stage 1, we use the English retrieval fine-tuning data\footnote{\url{https://huggingface.co/datasets/lightonai/embeddings-fine-tuning-filtered-en}} released as part of the DenseOn and LateOn work~\citep{lighton}. This release contains retrieval training data with mined hard negatives filtered using the NV-Retriever~\citep{moreira2024nv} procedure, including English examples from the MLDR training split. The source data comprise 1.57M training records, each containing a query, one positive document, and ten negative documents. Since this stage minimizes the discrepancy between student and teacher relevance scores for individual query--document pairs, we convert each record into separate pairs by pairing the query with the positive document and the first three negatives in the provided candidate order. We retain all records, yielding 6.28M training pairs. For Query-Local Relevance Discrimination in Stage 2, we use RLHN-100K\footnote{\url{https://huggingface.co/datasets/rlhn/rlhn-100K}}~\citep{rlhn}, which provides positive and negative documents for each query. It is derived from seven datasets in the BGE training collection~\citep{li2025making}, with cascading LLMs used to identify and relabel false negatives while leaving the ArguAna subset unchanged. This stage learns relative relevance scores within the same query, so we organize the provided documents into candidate sets. We use all records and retain the first positive document and the first seven negatives in the provided order for each record. Each query and its candidate set form a training instance, with contrastive comparisons restricted to that set.

\paragraph{Hyperparams}
Both stages are trained for one epoch using AdamW~\citep{adamw} with a learning rate of $5\times10^{-5}$, no weight decay, and a linear learning rate scheduler. The warmup ratios are 0.10 and 0.05 for Stage 1 and Stage 2, respectively. We clip the gradient norm to 1.0 and enable BF16 mixed precision and gradient checkpointing. Stage 1 uses 2 pairs per GPU with 64 gradient accumulation steps, resulting in a total batch size of 1,024 pairs. Stage 2 uses 8 pairs per GPU with 2 gradient accumulation steps, resulting in a total batch size of 128 pairs. Each query is paired with one positive and seven negative documents. The temperature for the Stage 2 contrastive loss is set to 0.1. The maximum sequence length is set to 32,768 for both training stages and inference. Both stages apply LoRA to the text decoder with a rank of 64, an alpha of 128, and a dropout rate of 0.05, while keeping the vision encoder and visual feature mergers frozen.

\paragraph{Reproducibility}
We implement both training stages using Sentence Transformers\footnote{\url{https://github.com/huggingface/sentence-transformers}}~\citep{sbert} and set the random seed to 42 for training, including data shuffling. The reported results are obtained from a single training run for each configuration. Data preparation procedures and training hyperparameters are specified above to facilitate reproduction of the experiments.

\paragraph{Hardware}
We trained RenderRank on eight NVIDIA RTX A6000 GPUs, each with 48GB of memory, using a server equipped with two Intel Xeon Gold 6230R processors, each with 26 cores. Inference throughput was measured on a single GPU.

\section{Detailed Results}

We provide the full results for each of the 11 BEIR datasets. Tables~\ref{tab:beir_main_tokens} and~\ref{tab:beir_main_pps} report the average number of input tokens and inference throughput for each evaluated reranker. Tables~\ref{tab:beir_font_performance}, \ref{tab:beir_font_tokens}, and~\ref{tab:beir_font_pps} present the reranking performance, average number of input tokens, and throughput of RenderRank across rendering font sizes. Table~\ref{tab:beir_ablation} provides the full performance results for the training-stage ablations.

\input{Aappendix_Tables/main_full}

\input{Aappendix_Tables/font_variation}

\input{Aappendix_Tables/ablation_variation}

%% file: Aappendix_Tables/App_gen.tex
\definecolor{renderblue}{HTML}{2196D2}
\definecolor{renderred}{HTML}{E57373}

\begin{table*}[b]
\centering
\caption{Performance and document-token counts across rendering configurations. QASPER is evaluated using F1, and GovReport and Multi-News using ROUGE-L. Tokens denotes the average document-token count. Blue shading indicates a reduction in document-token count relative to text input, while red shading indicates an increase.}
\label{tab:rendering_config_full}
\small
\setlength{\tabcolsep}{4pt}
\renewcommand{\arraystretch}{1.15}
\resizebox{0.85\textwidth}{!}{%
\begin{tabular}{*{14}{c}}
  \toprule
  \multirow{3}{*}{Font}
  & \multirow{3}{*}{LS}
  & \multicolumn{6}{c}{Gemma-4-26B-A4B-it}
  & \multicolumn{6}{c}{Qwen3.6-35B-A3B} \\
  \cmidrule(lr){3-8}\cmidrule(lr){9-14}
  & & \multicolumn{2}{c}{QASPER}
    & \multicolumn{2}{c}{GovReport}
    & \multicolumn{2}{c}{Multi-News}
    & \multicolumn{2}{c}{QASPER}
    & \multicolumn{2}{c}{GovReport}
    & \multicolumn{2}{c}{Multi-News} \\
  \cmidrule(lr){3-4}\cmidrule(lr){5-6}
  \cmidrule(lr){7-8}\cmidrule(lr){9-10}
  \cmidrule(lr){11-12}\cmidrule(lr){13-14}
  & & F1 & Tokens
    & ROUGE-L & Tokens
    & ROUGE-L & Tokens
    & F1 & Tokens
    & ROUGE-L & Tokens
    & ROUGE-L & Tokens \\
  \midrule
  Text & --
    & 47.73 & 4,998
    & 28.78 & 10,551
    & 22.57 & 2,748
    & 51.45 & 5,023
    & 31.10 & 10,579
    & 23.41 & 2,652 \\
  \midrule
  \multirow{3}{*}{10pt}
    & 1.0
    & 45.90 & \cellcolor{renderblue!19.35}3,064
    & 28.27 & \cellcolor{renderblue!20.51}6,223
    & 22.24 & \cellcolor{renderblue!16.46}1,843
    & 41.32 & \cellcolor{renderblue!30.98}1,911
    & 30.55 & \cellcolor{renderblue!30.16}4,197
    & 22.29 & \cellcolor{renderblue!30.21}1,050 \\
    & 1.1
    & 45.72 & \cellcolor{renderblue!15.65}3,434
    & 28.46 & \cellcolor{renderblue!16.40}7,090
    & 22.02 & \cellcolor{renderblue!12.77}2,046
    & 43.18 & \cellcolor{renderblue!28.18}2,192
    & 30.75 & \cellcolor{renderblue!27.20}4,823
    & 22.80 & \cellcolor{renderblue!27.33}1,202 \\
    & 1.2
    & 44.58 & \cellcolor{renderblue!14.33}3,566
    & 28.27 & \cellcolor{renderblue!14.70}7,449
    & 22.21 & \cellcolor{renderblue!11.79}2,100
    & 45.80 & \cellcolor{renderblue!27.08}2,303
    & 30.42 & \cellcolor{renderblue!25.99}5,079
    & 22.74 & \cellcolor{renderblue!26.37}1,253 \\
  \midrule
  \multirow{3}{*}{11pt}
    & 1.0
    & 47.48 & \cellcolor{renderblue!11.32}3,867
    & 28.53 & \cellcolor{renderblue!11.53}8,117
    & 22.40 & \cellcolor{renderblue!8.63}2,274
    & 45.26 & \cellcolor{renderblue!24.84}2,528
    & 30.53 & \cellcolor{renderblue!23.67}5,571
    & 22.59 & \cellcolor{renderblue!24.02}1,378 \\
    & 1.1
    & 48.32 & \cellcolor{renderblue!6.38}4,361
    & 28.69 & \cellcolor{renderblue!6.76}9,125
    & 22.24 & \cellcolor{renderblue!3.20}2,572
    & 46.46 & \cellcolor{renderblue!21.67}2,846
    & 30.55 & \cellcolor{renderblue!20.21}6,303
    & 23.12 & \cellcolor{renderblue!21.07}1,535 \\
    & 1.2
    & 45.76 & \cellcolor{renderblue!4.19}4,579
    & 28.68 & \cellcolor{renderblue!4.05}9,696
    & 22.14 & \cellcolor{renderblue!1.03}2,691
    & 47.41 & \cellcolor{renderblue!19.95}3,019
    & 30.85 & \cellcolor{renderblue!18.44}6,678
    & 23.19 & \cellcolor{renderblue!19.15}1,637 \\
  \midrule
  \multirow{3}{*}{12pt}
    & 1.0
    & 48.07 & \cellcolor{renderblue!6.59}4,339
    & 28.67 & \cellcolor{renderblue!7.01}9,072
    & 22.20 & \cellcolor{renderblue!3.79}2,539
    & 48.72 & \cellcolor{renderblue!21.83}2,829
    & 30.82 & \cellcolor{renderblue!20.40}6,262
    & 22.98 & \cellcolor{renderblue!21.33}1,521 \\
    & 1.1
    & 47.93 & \cellcolor{renderblue!1.50}4,848
    & 28.75 & \cellcolor{renderblue!1.35}10,266
    & 21.90 & \cellcolor{renderred!15.30}2,764
    & 45.35 & \cellcolor{renderblue!17.91}3,223
    & 30.66 & \cellcolor{renderblue!16.27}7,136
    & 23.35 & \cellcolor{renderblue!17.48}1,725 \\
    & 1.2
    & 46.50 & \cellcolor{renderred!17.76}5,274
    & 28.60 & \cellcolor{renderred!19.03}11,401
    & 22.24 & \cellcolor{renderred!20.74}3,063
    & 45.33 & \cellcolor{renderblue!14.59}3,557
    & 30.69 & \cellcolor{renderblue!12.46}7,943
    & 23.41 & \cellcolor{renderblue!13.96}1,912 \\
  \bottomrule
\end{tabular}%
}
\end{table*}

%% file: Aappendix_Tables/dataset.tex
\begin{table}[t]
  \centering
  \caption{Dataset composition for the two training stages.}
  \label{tab:training_data}

  \small
  \setlength{\tabcolsep}{13pt}
  \renewcommand{\arraystretch}{1.1}
  \resizebox{0.85\linewidth}{!}{%
  \begin{tabular}{lrrlr}
    \toprule

    \multicolumn{3}{c}{\textbf{Stage 1}}
    & \multicolumn{2}{c}{\textbf{Stage 2}} \\

    \cmidrule{1-3}
    \cmidrule{4-5}

    \textbf{Dataset}
    & \textbf{Records}
    & \textbf{Pairs}
    & \textbf{Dataset}
    & \textbf{Records} \\

    \midrule

    FiQA
    & 13,008
    & 52,032
    & ArguAna
    & 4,065 \\

    HotpotQA
    & 142,537
    & 570,148
    & FEVER
    & 8,151 \\

    Natural Questions
    & 114,030
    & 456,120
    & FiQA
    & 5,496 \\

    MS MARCO
    & 518,851
    & 2,075,404
    & HotpotQA
    & 10,246 \\

    FEVER
    & 126,852
    & 507,408
    & MS MARCO Passage
    & 46,926 \\

    SQuAD v2
    & 129,241
    & 516,964
    & Natural Questions
    & 6,043 \\

    TriviaQA
    & 511,074
    & 2,044,296
    & SciDocsRR
    & 12,654 \\

    MIRACL
    & 6,191
    & 24,764
    &
    & \\

    MLDR
    & 9,086
    & 36,344
    &
    & \\

    \midrule

    \textbf{Total}
    & \textbf{1,570,870}
    & \textbf{6,283,480}
    & \textbf{Total}
    & \textbf{93,581} \\

    \bottomrule
  \end{tabular}%
  }
\end{table}

%% file: Aappendix_Tables/main_full.tex
\begin{table*}[h]
    \centering
    \caption{Full results for average input token counts of rerankers on 11 BEIR datasets.}

    \label{tab:beir_main_tokens}
    \scriptsize
    \renewcommand{\arraystretch}{1.2}
    \setlength{\tabcolsep}{3.2pt}
    \resizebox{\textwidth}{!}{
    \begin{tabular}{lccccccccccccc}
    \toprule
    & & \multicolumn{12}{c}{\textbf{BEIR}} \\
    \cmidrule(lr){3-14}
    \textbf{Model} & \textbf{Params} & \textbf{AA} & \textbf{CFV} & \textbf{DBP} & \textbf{FQA} & \textbf{FVR} & \textbf{HQA} & \textbf{NFC} & \textbf{SD} & \textbf{SF} & \textbf{TC} & \textbf{TCH} & \textbf{Avg.} \\
    \midrule
    gte-reranker-modernbert-base & 150M & 590.82 & 349.43 & 97.31 & 388.50 & 224.47 & 125.54 & 357.15 & 287.82 & 365.39 & 345.06 & 820.29 & 359.25 \\
    mxbai-rerank-large-v1 & 435M & 563.76 & 329.37 & 89.29 & 383.63 & 205.76 & 116.99 & 342.54 & 276.43 & 354.01 & 341.52 & 818.69 & 347.45 \\
    bge-reranker-large & 560M & 642.82 & 417.03 & 104.19 & 420.74 & 255.46 & 134.94 & 420.11 & 339.02 & 448.81 & 408.58 & 910.87 & 409.32 \\
    Qwen3-Reranker-0.6B & 596M & 670.01 & 429.17 & 169.26 & 460.64 & 303.84 & 199.44 & 449.36 & 353.84 & 464.23 & 435.73 & 890.13 & 438.70 \\
    LAMAR-600m & 600M & 642.82 & 417.03 & 104.19 & 420.74 & 255.46 & 134.94 & 420.11 & 339.02 & 448.81 & 408.58 & 910.87 & 409.32 \\
    llama-nemotron-rerank-1b-v2 & 1.2B & 591.56 & 353.67 & 99.53 & 387.35 & 228.67 & 128.74 & 370.61 & 284.69 & 387.86 & 357.02 & 814.20 & 363.99 \\
    mxbai-rerank-large-v2 & 1.5B & 681.26 & 440.21 & 180.27 & 471.69 & 314.92 & 210.46 & 460.37 & 364.93 & 475.25 & 446.78 & 901.25 & 449.76 \\
    LightOn-rerank-PW-2B & 2.2B & 647.73 & 406.94 & 146.84 & 439.11 & 281.40 & 177.41 & 426.36 & 331.34 & 441.63 & 411.52 & 868.13 & 416.22 \\
    bge-reranker-v2-gemma & 2.5B & 631.43 & 388.00 & 133.51 & 432.04 & 264.31 & 163.39 & 401.57 & 316.95 & 410.03 & 389.95 & 859.04 & 399.11 \\
    Qwen3-Reranker-4B & 4B & 670.01 & 429.17 & 169.26 & 460.64 & 303.84 & 199.44 & 449.36 & 353.84 & 464.23 & 435.73 & 890.13 & 438.70 \\
    zerank-2-reranker & 4B & 612.11 & 371.28 & 111.10 & 402.19 & 246.37 & 141.67 & 391.17 & 295.13 & 406.51 & 377.76 & 832.26 & 380.69 \\
    LightOn-rerank-PW-4B & 4.5B & 647.73 & 406.94 & 146.84 & 439.11 & 281.40 & 177.41 & 426.36 & 331.34 & 441.63 & 411.52 & 868.13 & 416.22 \\
    \midrule
    \textbf{RenderRank (Ours)} & 2.1B & 518.35 & 283.33 & 128.82 & 280.62 & 197.20 & 150.52 & 278.68 & 257.07 & 292.15 & 284.13 & 519.92 & 290.07 \\
    \bottomrule
    \end{tabular}
    }
\end{table*}

\begin{table*}[h]
    \centering    
    \caption{Full throughput results (PPS) of rerankers on 11 BEIR datasets.}

    \label{tab:beir_main_pps}
    \scriptsize
    \renewcommand{\arraystretch}{1.2}
    \setlength{\tabcolsep}{3.5pt}
    \resizebox{\textwidth}{!}{
    \begin{tabular}{lccccccccccccc}
    \toprule
    & & \multicolumn{12}{c}{\textbf{BEIR}} \\
    \cmidrule(lr){3-14}
    \textbf{Model} & \textbf{Params} & \textbf{AA} & \textbf{CFV} & \textbf{DBP} & \textbf{FQA} & \textbf{FVR} & \textbf{HQA} & \textbf{NFC} & \textbf{SD} & \textbf{SF} & \textbf{TC} & \textbf{TCH} & \textbf{Avg.} \\
    \midrule
    gte-reranker-modernbert-base & 150M & 45.35 & 60.35 & 378.30 & 44.93 & 88.48 & 235.90 & 83.25 & 72.61 & 67.88 & 83.00 & 16.89 & 106.99 \\
    mxbai-rerank-large-v1 & 435M & 77.62 & 72.92 & 442.98 & 72.03 & 101.50 & 229.39 & 82.19 & 102.99 & 80.61 & 83.36 & 77.97 & 129.42 \\
    bge-reranker-large & 560M & 48.05 & 48.46 & 167.70 & 35.59 & 53.84 & 101.97 & 42.00 & 52.59 & 49.08 & 49.06 & 48.27 & 63.33 \\
    Qwen3-Reranker-0.6B & 596M & 32.87 & 41.48 & 183.17 & 37.38 & 60.97 & 127.17 & 55.55 & 56.16 & 51.67 & 55.43 & 18.70 & 65.50 \\
    LAMAR-600m & 600M & 31.45 & 32.28 & 202.17 & 32.72 & 75.28 & 150.04 & 48.65 & 58.68 & 44.76 & 49.08 & 17.98 & 67.55 \\
    llama-nemotron-rerank-1b-v2 & 1.24B & 40.33 & 52.24 & 259.71 & 42.13 & 72.90 & 163.05 & 64.35 & 64.48 & 59.70 & 64.75 & 21.09 & 82.25 \\
    mxbai-rerank-large-v2 & 1.5B & 35.20 & 50.82 & 133.88 & 49.34 & 70.29 & 94.71 & 48.11 & 56.99 & 47.71 & 45.03 & 24.95 & 59.73 \\
    LightOn-rerank-PW-2B & 2.2B & 15.38 & 22.57 & 66.08 & 23.53 & 34.34 & 53.68 & 20.50 & 31.10 & 22.40 & 24.18 & 12.72 & 29.68 \\
    bge-reranker-v2-gemma & 2.51B & 18.14 & 23.56 & 97.04 & 18.46 & 45.49 & 67.15 & 28.25 & 28.85 & 26.17 & 20.32 & 17.19 & 35.51 \\
    Qwen3-Reranker-4B & 4B & 12.59 & 18.62 & 50.42 & 17.60 & 25.75 & 40.33 & 17.14 & 23.35 & 14.91 & 18.51 & 7.33 & 22.41 \\
    zerank-2-reranker & 4.02B & 11.42 & 21.06 & 70.42 & 20.14 & 30.58 & 53.61 & 19.50 & 27.80 & 19.27 & 20.69 & 9.20 & 27.61 \\
    LightOn-rerank-PW-4B & 4.5B & 7.00 & 9.34 & 25.01 & 9.89 & 14.54 & 22.25 & 9.45 & 13.02 & 8.52 & 10.05 & 5.22 & 12.21 \\
    \midrule
    \textbf{RenderRank (Ours)} & 2.1B & 35.53 & 68.97 & 151.77 & 64.43 & 87.86 & 113.36 & 66.41 & 79.16 & 68.79 & 70.79 & 38.12 & 76.83 \\
    \bottomrule
    \end{tabular}
    }
\end{table*}

%% file: Aappendix_Tables/font_variation.tex
\begin{table*}[h]
    \centering 
    \caption{Full reranking performance results (NDCG@10) for RenderRank with different rendering font sizes on BEIR.}

    \label{tab:beir_font_performance}
    \scriptsize
    \renewcommand{\arraystretch}{1.1}
    \setlength{\tabcolsep}{4pt}
    \resizebox{0.85\textwidth}{!}{
    \begin{tabular}{lcccccccccccc}
    \toprule
    & \multicolumn{12}{c}{\textbf{BEIR}} \\
    \cmidrule(lr){2-13}
    \textbf{Input} & \textbf{AA} & \textbf{CFV} & \textbf{DBP} & \textbf{FQA} & \textbf{FVR} & \textbf{HQA} & \textbf{NFC} & \textbf{SD} & \textbf{SF} & \textbf{TC} & \textbf{TCH} & \textbf{Avg.} \\
    \midrule
    10\,pt & 67.42 & 33.79 & 45.43 & 40.83 & 88.20 & 77.94 & 36.26 & 20.09 & 75.97 & 83.23 & 36.24 & 55.04 \\
    11\,pt & 68.41 & 34.42 & 45.97 & 41.92 & 88.87 & 78.68 & 36.49 & 20.69 & 76.18 & 83.73 & 35.16 & 55.50 \\
    12\,pt & 69.11 & 35.91 & 46.19 & 42.66 & 88.92 & 78.60 & 37.31 & 20.75 & 77.00 & 84.82 & 34.27 & 55.96 \\
    13\,pt & 68.82 & 35.68 & 46.38 & 42.44 & 89.31 & 79.05 & 36.99 & 20.97 & 77.82 & 84.31 & 34.46 & 56.02 \\
    14\,pt & 69.54 & 36.24 & 47.32 & 42.92 & 89.30 & 79.17 & 37.30 & 21.09 & 78.08 & 85.06 & 34.67 & 56.43 \\
    \bottomrule
    \end{tabular}
    }
\end{table*}

\begin{table*}[h]
    \centering
    \caption{Full results for average input token counts of RenderRank with different rendering font sizes on BEIR.}
    \label{tab:beir_font_tokens}
    \scriptsize
    \renewcommand{\arraystretch}{1.15}
    \setlength{\tabcolsep}{3.8pt}
    \resizebox{0.95\textwidth}{!}{
    \begin{tabular}{lcccccccccccc}
    \toprule
    & \multicolumn{12}{c}{\textbf{BEIR}} \\
    \cmidrule(lr){2-13}
    \textbf{Input} & \textbf{AA} & \textbf{CFV} & \textbf{DBP} & \textbf{FQA} & \textbf{FVR} & \textbf{HQA} & \textbf{NFC} & \textbf{SD} & \textbf{SF} & \textbf{TC} & \textbf{TCH} & \textbf{Avg.} \\
    \midrule
    10\,pt & 460.10 & 228.64 & 121.91 & 223.08 & 167.75 & 139.79 & 218.55 & 207.48 & 231.83 & 225.19 & 376.91 & 236.47 \\
    11\,pt & 502.14 & 267.68 & 124.87 & 264.16 & 189.13 & 147.21 & 263.43 & 244.60 & 276.35 & 268.96 & 474.97 & 274.86 \\
    12\,pt & 518.35 & 283.33 & 128.82 & 280.62 & 197.20 & 150.52 & 278.68 & 257.07 & 292.15 & 284.13 & 519.92 & 290.07 \\
    13\,pt & 549.75 & 313.62 & 147.08 & 312.76 & 219.96 & 168.11 & 310.62 & 284.64 & 323.66 & 315.89 & 586.41 & 321.14 \\
    14\,pt & 605.08 & 364.90 & 151.26 & 366.89 & 248.68 & 178.39 & 367.26 & 333.28 & 380.95 & 371.74 & 709.15 & 370.69 \\
    \bottomrule
    \end{tabular}
    }
\end{table*}

\begin{table*}[h]
    \centering
    \caption{Full throughput results (PPS) for RenderRank with different rendering font sizes on BEIR.}

    \label{tab:beir_font_pps}
    \scriptsize
    \renewcommand{\arraystretch}{1.1}
    \setlength{\tabcolsep}{4pt}
    \resizebox{0.85\textwidth}{!}{
    \begin{tabular}{lcccccccccccc}
    \toprule
    & \multicolumn{12}{c}{\textbf{BEIR}} \\
    \cmidrule(lr){2-13}
    \textbf{Input} & \textbf{AA} & \textbf{CFV} & \textbf{DBP} & \textbf{FQA} & \textbf{FVR} & \textbf{HQA} & \textbf{NFC} & \textbf{SD} & \textbf{SF} & \textbf{TC} & \textbf{TCH} & \textbf{Avg.} \\
    \midrule
    10\,pt & 37.89 & 74.65 & 167.24 & 89.46 & 116.13 & 133.59 & 93.31 & 96.62 & 86.20 & 89.76 & 54.18 & 94.46 \\
    11\,pt & 35.94 & 72.68 & 159.66 & 75.50 & 101.63 & 115.95 & 76.49 & 83.03 & 72.76 & 67.07 & 42.30 & 82.09 \\
    12\,pt & 35.53 & 68.97 & 151.77 & 64.43 & 87.86 & 113.36 & 66.41 & 79.16 & 68.79 & 70.79 & 38.12 & 76.83 \\
    13\,pt & 33.60 & 62.36 & 126.00 & 64.40 & 88.58 & 113.85 & 66.10 & 72.04 & 62.47 & 64.50 & 30.08 & 71.27 \\
    14\,pt & 31.70 & 47.73 & 134.06 & 55.04 & 70.06 & 107.84 & 56.17 & 61.72 & 47.09 & 55.35 & 27.33 & 63.10 \\
    \bottomrule
    \end{tabular}
    }
\end{table*}

%% file: Aappendix_Tables/ablation_variation.tex
\begin{table*}[t]
    \centering
    \caption{Full training-stage ablation results (NDCG@10) with text and image inputs on 11 BEIR datasets.}

    \label{tab:beir_ablation}
    \scriptsize
    \renewcommand{\arraystretch}{1.1}
    \setlength{\tabcolsep}{3pt}
    \resizebox{0.93\textwidth}{!}{
    \begin{tabular}{llcccccccccccc}
    \toprule
    & & \multicolumn{12}{c}{\textbf{BEIR}} \\
    \cmidrule(lr){3-14}
    \textbf{Model} & \textbf{Input} & \textbf{AA} & \textbf{CFV} & \textbf{DBP} & \textbf{FQA} & \textbf{FVR} & \textbf{HQA} & \textbf{NFC} & \textbf{SD} & \textbf{SF} & \textbf{TC} & \textbf{TCH} & \textbf{Avg.} \\
    \midrule
    Qwen3-VL-Reranker-2B & Text & 57.31 & 32.10 & 45.33 & 43.22 & 86.85 & 77.86 & 37.47 & 20.71 & 77.39 & 82.87 & 35.17 & 54.21 \\
     & Image & 58.68 & 22.92 & 40.84 & 36.32 & 79.57 & 73.23 & 35.27 & 18.84 & 77.67 & 79.13 & 29.69 & 50.20 \\
    \midrule
    Stage 1 & Text & 57.06 & 32.68 & 46.43 & 44.37 & 88.88 & 77.88 & 37.67 & 20.99 & 78.32 & 84.67 & 40.75 & 55.43 \\
     & Image & 60.56 & 36.17 & 45.49 & 41.77 & 89.00 & 78.19 & 36.72 & 20.14 & 76.83 & 83.47 & 35.15 & 54.86 \\
    \midrule
    Stage 2 (RenderRank) & Text & 71.23 & 33.34 & 46.76 & 45.01 & 89.10 & 78.74 & 37.85 & 22.18 & 78.42 & 85.34 & 41.37 & 57.21 \\
     & Image & 69.11 & 35.91 & 46.19 & 42.66 & 88.92 & 78.60 & 37.31 & 20.75 & 77.00 & 84.82 & 34.27 & 55.96 \\
    \bottomrule
    \end{tabular}
    }
\end{table*}

%% file: iclr2027_conference.bib
@article{lewis2020retrieval,
  title={Retrieval-augmented generation for knowledge-intensive nlp tasks},
  author={Lewis, Patrick and Perez, Ethan and Piktus, Aleksandra and Petroni, Fabio and Karpukhin, Vladimir and Goyal, Naman and K{\"u}ttler, Heinrich and Lewis, Mike and Yih, Wen-tau and Rockt{\"a}schel, Tim and others},
  journal={Advances in neural information processing systems},
  volume={33},
  pages={9459--9474},
  year={2020}
}

@inproceedings{izacard2021leveraging,
  title={Leveraging passage retrieval with generative models for open domain question answering},
  author={Izacard, Gautier and Grave, Edouard},
  booktitle={Proceedings of the 16th conference of the european chapter of the association for computational linguistics: main volume},
  pages={874--880},
  year={2021}
}

@article{izacard2023atlas,
  title={Atlas: Few-shot learning with retrieval augmented language models},
  author={Izacard, Gautier and Lewis, Patrick and Lomeli, Maria and Hosseini, Lucas and Petroni, Fabio and Schick, Timo and Dwivedi-Yu, Jane and Joulin, Armand and Riedel, Sebastian and Grave, Edouard},
  journal={Journal of Machine Learning Research},
  volume={24},
  number={251},
  pages={1--43},
  year={2023}
}

@inproceedings{karpukhin2020dense,
  title={Dense passage retrieval for open-domain question answering},
  author={Karpukhin, Vladimir and Oguz, Barlas and Min, Sewon and Lewis, Patrick and Wu, Ledell and Edunov, Sergey and Chen, Danqi and Yih, Wen-tau},
  booktitle={Proceedings of the 2020 conference on empirical methods in natural language processing (EMNLP)},
  pages={6769--6781},
  year={2020}
}

@article{xiong2020approximate,
  title={Approximate nearest neighbor negative contrastive learning for dense text retrieval},
  author={Xiong, Lee and Xiong, Chenyan and Li, Ye and Tang, Kwok-Fung and Liu, Jialin and Bennett, Paul and Ahmed, Junaid and Overwijk, Arnold},
  journal={arXiv preprint arXiv:2007.00808},
  year={2020}
}

@inproceedings{khattab2020colbert,
  title={Colbert: Efficient and effective passage search via contextualized late interaction over bert},
  author={Khattab, Omar and Zaharia, Matei},
  booktitle={Proceedings of the 43rd International ACM SIGIR conference on research and development in Information Retrieval},
  pages={39--48},
  year={2020}
}

@article{yu2024rankrag,
  title={Rankrag: Unifying context ranking with retrieval-augmented generation in llms},
  author={Yu, Yue and Ping, Wei and Liu, Zihan and Wang, Boxin and You, Jiaxuan and Zhang, Chao and Shoeybi, Mohammad and Catanzaro, Bryan},
  journal={Advances in neural information processing systems},
  volume={37},
  pages={121156--121184},
  year={2024}
}

@inproceedings{asai2024self,
  title={Self-rag: Learning to retrieve, generate, and critique through self-reflection},
  author={Asai, Akari and Wu, Zeqiu and Wang, Yizhong and Sil, Avi and Hajishirzi, Hannaneh},
  booktitle={International conference on learning representations},
  volume={2024},
  pages={9112--9141},
  year={2024}
}

@inproceedings{glass2022re2g,
  title={Re2G: Retrieve, rerank, generate},
  author={Glass, Michael and Rossiello, Gaetano and Chowdhury, Md Faisal Mahbub and Naik, Ankita Rajaram and Cai, Pengshan and Gliozzo, Alfio},
  booktitle={Proceedings of the 2022 Conference of the North American Chapter of the Association for Computational Linguistics: Human Language Technologies},
  pages={2701--2715},
  year={2022}
}

@article{li2023parade,
  title={Parade: Passage representation aggregation fordocument reranking},
  author={Li, Canjia and Yates, Andrew and MacAvaney, Sean and He, Ben and Sun, Yingfei},
  journal={ACM Transactions on Information Systems},
  volume={42},
  number={2},
  pages={1--26},
  year={2023},
  publisher={ACM New York, NY, USA}
}

@article{wang2024qwen2,
  title={Qwen2-vl: Enhancing vision-language model's perception of the world at any resolution},
  author={Wang, Peng and Bai, Shuai and Tan, Sinan and Wang, Shijie and Fan, Zhihao and Bai, Jinze and Chen, Keqin and Liu, Xuejing and Wang, Jialin and Ge, Wenbin and others},
  journal={arXiv preprint arXiv:2409.12191},
  year={2024}
}

@article{rust2022language,
  title={Language modelling with pixels},
  author={Rust, Phillip and Lotz, Jonas F and Bugliarello, Emanuele and Salesky, Elizabeth and de Lhoneux, Miryam and Elliott, Desmond},
  journal={arXiv preprint arXiv:2207.06991},
  year={2022}
}

@article{lyu2025pixelworld,
  title={PixelWorld: How Far Are We from Perceiving Everything as Pixels?},
  author={Lyu, Zhiheng and Ma, Xueguang and Chen, Wenhu},
  journal={arXiv preprint arXiv:2501.19339},
  year={2025}
}

@inproceedings{li2025text,
  title={Text or Pixels? Evaluating Efficiency and Understanding of LLMs with Visual Text Inputs.},
  author={Li, Yanhong and Lan, Zixuan and Zhou, Jiawei},
  booktitle={EMNLP (Findings)},
  pages={10564--10578},
  year={2025}
}

@article{tang2026visual,
  title={Visual Text Compression as Measure Transport},
  author={Tang, Lv and Zheng, Tianyi and Liu, Yang and Li, Bo and Li, Xingyu},
  journal={arXiv preprint arXiv:2605.06708},
  year={2026}
}

@article{yuan2026design,
  title={On the Design Fundamentals of Pixel Text Representation Learning},
  author={Yuan, Chaohao and Yuan, Ruifeng and Huang, Zhuoxu and Rong, Yu and Cheng, Hong and Chan, Hou Pong and Xiao, Chenghao},
  journal={arXiv preprint arXiv:2609.01147},
  year={2026}
}

@inproceedings{tschannen2023clippo,
  title={Clippo: Image-and-language understanding from pixels only},
  author={Tschannen, Michael and Mustafa, Basil and Houlsby, Neil},
  booktitle={2023 IEEE/CVF Conference on Computer Vision and Pattern Recognition (CVPR)},
  pages={11006--11017},
  year={2023},
  organization={IEEE}
}

@article{xiao2024pixel,
  title={Pixel sentence representation learning},
  author={Xiao, Chenghao and Huang, Zhuoxu and Chen, Danlu and Hudson, G Thomas and Li, Yizhi and Duan, Haoran and Lin, Chenghua and Fu, Jie and Han, Jungong and Moubayed, Noura Al},
  journal={arXiv preprint arXiv:2402.08183},
  year={2024}
}

@inproceedings{lotz2023text,
  title={Text rendering strategies for pixel language models},
  author={Lotz, Jonas F and Salesky, Elizabeth and Rust, Phillip and Elliott, Desmond},
  booktitle={Proceedings of the 2023 Conference on Empirical Methods in Natural Language Processing},
  pages={10155--10172},
  year={2023}
}

@inproceedings{cheng2026glyph,
  title={Glyph: Scaling context windows via visual-text compression},
  author={Cheng, Jiale and Liu, Yusen and Zhang, Xinyu and Fei, Yulin and Hong, Wenyi and Lyu, Ruiliang and Wang, Weihan and Su, Zhe and Gu, Xiaotao and Liu, Xiao and others},
  booktitle={Proceedings of the 64th Annual Meeting of the Association for Computational Linguistics (Volume 1: Long Papers)},
  pages={37145--37158},
  year={2026}
}

@inproceedings{
taraday2025efficient,
title={Efficient Discriminative Joint Encoders for Large Scale Vision-Language Reranking},
author={Mitchell Keren Taraday and Shahaf Wagner and Chaim Baskin},
booktitle={The Fourteenth International Conference on Learning Representations},
year={2026},
url={https://openreview.net/forum?id=UXtTBAyqVB}
}

@article{sun2026very,
  title={Very efficient listwise multimodal reranking for long documents},
  author={Sun, Yiqun and Wei, Pengfei and Hsieh, Lawrence B},
  journal={arXiv preprint arXiv:2605.11864},
  year={2026}
}

@article{dejean2026efficient,
  title={Efficient Listwise Reranking with Compressed Document Representations},
  author={D{\'e}jean, Herv{\'e} and Clinchant, St{\'e}phane},
  journal={arXiv preprint arXiv:2604.26483},
  year={2026}
}

@article{lu2026comprank,
  title={CompRank: Efficient LLM Reranking via Token-Level Compression and Decoding-Free Scoring},
  author={Lu, Xuan and Huang, Haohang and Fan, Yingqi and Tong, Junlong and Zhang, Yuxuan and Nie, Ping and Meng, Rui and Shen, Xiaoyu},
  journal={arXiv preprint arXiv:2606.11700},
  year={2026}
}

@article{zhuang2026layer,
  title={Layer-wise Token Compression for Efficient Document Reranking},
  author={Zhuang, Shengyao and Xu, Zhichao and Lauriola, Ivano},
  journal={arXiv preprint arXiv:2605.20683},
  year={2026}
}

@article{wang2026tevatron,
  title={Tevatron-Elastic: A Unified Abstraction for Training Elastic Retrievers and Rerankers},
  author={Wang, Yu and Zhuang, Shengyao and Ma, Xueguang and Wu, Zongyu and Lin, Jimmy and Srikumar, Vivek and Xu, Zhichao},
  journal={arXiv preprint arXiv:2608.08809},
  year={2026}
}

@article{nogueira2019passage,
  title={Passage Re-ranking with BERT},
  author={Nogueira, Rodrigo and Cho, Kyunghyun},
  journal={arXiv preprint arXiv:1901.04085},
  year={2019}
}

@inproceedings{nogueira2020document,
  title={Document ranking with a pretrained sequence-to-sequence model},
  author={Nogueira, Rodrigo and Jiang, Zhiying and Pradeep, Ronak and Lin, Jimmy},
  booktitle={Findings of the association for computational linguistics: EMNLP 2020},
  pages={708--718},
  year={2020}
}

@article{hofstatter2020improving,
  title={Improving efficient neural ranking models with cross-architecture knowledge distillation},
  author={Hofst{\"a}tter, Sebastian and Althammer, Sophia and Schr{\"o}der, Michael and Sertkan, Mete and Hanbury, Allan},
  journal={arXiv preprint arXiv:2010.02666},
  year={2020}
}

@inproceedings{baldelli2024twolar,
  title={TWOLAR: a TWO-step LLM-Augmented distillation method for passage Reranking},
  author={Baldelli, Davide and Jiang, Junfeng and Aizawa, Akiko and Torroni, Paolo},
  booktitle={European Conference on Information Retrieval},
  pages={470--485},
  year={2024},
  organization={Springer}
}

@inproceedings{schlatt2025rank,
  title={Rank-distillm: Closing the effectiveness gap between cross-encoders and llms for passage re-ranking},
  author={Schlatt, Ferdinand and Fr{\"o}be, Maik and Scells, Harrisen and Zhuang, Shengyao and Koopman, Bevan and Zuccon, Guido and Stein, Benno and Potthast, Martin and Hagen, Matthias},
  booktitle={European Conference on Information Retrieval},
  pages={323--334},
  year={2025},
  organization={Springer}
}

@inproceedings{sun2023chatgpt,
  title={Is ChatGPT good at search? investigating large language models as re-ranking agents},
  author={Sun, Weiwei and Yan, Lingyong and Ma, Xinyu and Wang, Shuaiqiang and Ren, Pengjie and Chen, Zhumin and Yin, Dawei and Ren, Zhaochun},
  booktitle={Proceedings of the 2023 conference on empirical methods in natural language processing},
  pages={14918--14937},
  year={2023}
}

@inproceedings{ma2024fine,
  title={Fine-tuning llama for multi-stage text retrieval},
  author={Ma, Xueguang and Wang, Liang and Yang, Nan and Wei, Furu and Lin, Jimmy},
  booktitle={Proceedings of the 47th International ACM SIGIR Conference on Research and Development in Information Retrieval},
  pages={2421--2425},
  year={2024}
}

@article{qwen3vlembedding,
  title={Qwen3-VL-Embedding and Qwen3-VL-Reranker: A Unified Framework for State-of-the-Art Multimodal Retrieval and Ranking},
  author={Li, Mingxin and Zhang, Yanzhao and Long, Dingkun and Chen Keqin and Song, Sibo and Bai, Shuai and Yang, Zhibo and Xie, Pengjun and Yang, An and Liu, Dayiheng and Zhou, Jingren and Lin, Junyang},
  journal={arXiv preprint arXiv:2601.04720},
  year={2026}
}

@misc{sourty2026denseonlateonfullyopen,
  title         = {DenseOn with the LateOn: Fully Open Dense and Late-Interaction Models for Multilingual, Long-Context, and Code Search},
  author        = {Raphaël Sourty and Antoine Chaffin and Paulo Roberto Moura Junior and Amélie Chatelain},
  year          = {2026},
  eprint        = {2607.27178},
  archivePrefix = {arXiv},
  primaryClass  = {cs.CL},
  url           = {https://arxiv.org/abs/2607.27178},
}

@misc{rlhn,
      title={Fixing Data That Hurts Performance: Cascading LLMs to Relabel Hard Negatives for Robust Information Retrieval}, 
      author={Nandan Thakur and Crystina Zhang and Xueguang Ma and Jimmy Lin},
      year={2025},
      eprint={2505.16967},
      archivePrefix={arXiv},
      primaryClass={cs.IR},
      url={https://arxiv.org/abs/2505.16967}, 
}

@article{lora,
  title={Lora: Low-rank adaptation of large language models},
  author={Hu, Edward J and Shen, Yelong and Wallis, Phillip and Allen-Zhu, Zeyuan and Li, Yuanzhi and Wang, Shean and Wang, Lu and Chen, Weizhu},
  journal={arXiv preprint arXiv:2106.09685},
  year={2021}
}

@article{relora,
  title={Relora: High-rank training through low-rank updates, 2023},
  author={Lialin, Vladislav and Shivagunde, Namrata and Muckatira, Sherin and Rumshisky, Anna},
  journal={URL https://arxiv. org/abs/2307.05695},
  year={2023}
}

@article{beir,
  title={Beir: A heterogenous benchmark for zero-shot evaluation of information retrieval models},
  author={Thakur, Nandan and Reimers, Nils and R{\"u}ckl{\'e}, Andreas and Srivastava, Abhishek and Gurevych, Iryna},
  journal={arXiv preprint arXiv:2104.08663},
  year={2021}
}

@inproceedings{AA,
  title={Retrieval of the best counterargument without prior topic knowledge},
  author={Wachsmuth, Henning and Syed, Shahbaz and Stein, Benno},
  booktitle={Proceedings of the 56th Annual Meeting of the Association for Computational Linguistics (Volume 1: Long Papers)},
  pages={241--251},
  year={2018}
}

@article{CFV,
  title={Climate-fever: A dataset for verification of real-world climate claims},
  author={Diggelmann, Thomas and Boyd-Graber, Jordan and Bulian, Jannis and Ciaramita, Massimiliano and Leippold, Markus},
  journal={arXiv preprint arXiv:2012.00614},
  year={2020}
}

@inproceedings{DBP,
  title={DBpedia-entity v2: a test collection for entity search},
  author={Hasibi, Faegheh and Nikolaev, Fedor and Xiong, Chenyan and Balog, Krisztian and Bratsberg, Svein Erik and Kotov, Alexander and Callan, Jamie},
  booktitle={Proceedings of the 40th International ACM SIGIR Conference on Research and Development in Information Retrieval},
  pages={1265--1268},
  year={2017}
}

@article{FQA,
  title={Www'18 open challenge: financial opinion mining and question answering},
  author={Maia, Macedo and Handschuh, Siegfried and Freitas, Andr{\'e} and Davis, Brian and McDermott, Ross and Zarrouk, Manel and Balahur, Alexandra},
  year={2018},
  publisher={Association for Computing Machinery}
}

@inproceedings{FVR,
  title={FEVER: a large-scale dataset for fact extraction and VERification},
  author={Thorne, James and Vlachos, Andreas and Christodoulopoulos, Christos and Mittal, Arpit},
  booktitle={Proceedings of the 2018 Conference of the North American Chapter of the Association for Computational Linguistics: Human Language Technologies, Volume 1 (Long Papers)},
  pages={809--819},
  year={2018}
}

@inproceedings{HQA,
  title={HotpotQA: A dataset for diverse, explainable multi-hop question answering},
  author={Yang, Zhilin and Qi, Peng and Zhang, Saizheng and Bengio, Yoshua and Cohen, William and Salakhutdinov, Ruslan and Manning, Christopher D},
  booktitle={Proceedings of the 2018 conference on empirical methods in natural language processing},
  pages={2369--2380},
  year={2018}
}

@inproceedings{NFC,
  title={A full-text learning to rank dataset for medical information retrieval},
  author={Boteva, Vera and Gholipour, Demian and Sokolov, Artem and Riezler, Stefan},
  booktitle={European Conference on Information Retrieval},
  pages={716--722},
  year={2016},
  organization={Springer}
}

@inproceedings{SD,
  title={Specter: Document-level representation learning using citation-informed transformers},
  author={Cohan, Arman and Feldman, Sergey and Beltagy, Iz and Downey, Doug and Weld, Daniel S},
  booktitle={Proceedings of the 58th annual meeting of the association for computational linguistics},
  pages={2270--2282},
  year={2020}
}

@inproceedings{SF,
  title={Fact or fiction: Verifying scientific claims},
  author={Wadden, David and Lin, Shanchuan and Lo, Kyle and Wang, Lucy Lu and van Zuylen, Madeleine and Cohan, Arman and Hajishirzi, Hannaneh},
  booktitle={Proceedings of the 2020 conference on empirical methods in natural language processing (EMNLP)},
  pages={7534--7550},
  year={2020}
}

@inproceedings{TC,
  title={TREC-COVID: constructing a pandemic information retrieval test collection},
  author={Voorhees, Ellen and Alam, Tasmeer and Bedrick, Steven and Demner-Fushman, Dina and Hersh, William R and Lo, Kyle and Roberts, Kirk and Soboroff, Ian and Wang, Lucy Lu},
  booktitle={ACM SIGIR Forum},
  volume={54},
  number={1},
  pages={1--12},
  year={2021},
  organization={ACM New York, NY, USA}
}

@inproceedings{TCH,
  title={Overview of touch{\'e} 2022: argument retrieval},
  author={Bondarenko, Alexander and Fr{\"o}be, Maik and Kiesel, Johannes and Syed, Shahbaz and Gurcke, Timon and Beloucif, Meriem and Panchenko, Alexander and Biemann, Chris and Stein, Benno and Wachsmuth, Henning and others},
  booktitle={International conference of the cross-language evaluation forum for European languages},
  pages={311--336},
  year={2022},
  organization={Springer}
}

@misc{bge,
      title={BGE M3-Embedding: Multi-Lingual, Multi-Functionality, Multi-Granularity Text Embeddings Through Self-Knowledge Distillation}, 
      author={Jianlv Chen and Shitao Xiao and Peitian Zhang and Kun Luo and Defu Lian and Zheng Liu},
      year={2024},
      eprint={2402.03216},
      archivePrefix={arXiv},
      primaryClass={cs.CL}
}

@misc{mmteb,
      title={MMTEB: Massive Multilingual Text Embedding Benchmark}, 
      author={Kenneth Enevoldsen and Isaac Chung and Imene Kerboua and Márton Kardos and Ashwin Mathur and David Stap and Jay Gala and Wissam Siblini and Dominik Krzemiński and Genta Indra Winata and Saba Sturua and Saiteja Utpala and Mathieu Ciancone and Marion Schaeffer and Gabriel Sequeira and Diganta Misra and Shreeya Dhakal and Jonathan Rystrøm and Roman Solomatin and Ömer Çağatan and Akash Kundu and Martin Bernstorff and Shitao Xiao and Akshita Sukhlecha and Bhavish Pahwa and Rafał Poświata and Kranthi Kiran GV and Shawon Ashraf and Daniel Auras and Björn Plüster and Jan Philipp Harries and Loïc Magne and Isabelle Mohr and Mariya Hendriksen and Dawei Zhu and Hippolyte Gisserot-Boukhlef and Tom Aarsen and Jan Kostkan and Konrad Wojtasik and Taemin Lee and Marek Šuppa and Crystina Zhang and Roberta Rocca and Mohammed Hamdy and Andrianos Michail and John Yang and Manuel Faysse and Aleksei Vatolin and Nandan Thakur and Manan Dey and Dipam Vasani and Pranjal Chitale and Simone Tedeschi and Nguyen Tai and Artem Snegirev and Michael Günther and Mengzhou Xia and Weijia Shi and Xing Han Lù and Jordan Clive and Gayatri Krishnakumar and Anna Maksimova and Silvan Wehrli and Maria Tikhonova and Henil Panchal and Aleksandr Abramov and Malte Ostendorff and Zheng Liu and Simon Clematide and Lester James Miranda and Alena Fenogenova and Guangyu Song and Ruqiya Bin Safi and Wen-Ding Li and Alessia Borghini and Federico Cassano and Hongjin Su and Jimmy Lin and Howard Yen and Lasse Hansen and Sara Hooker and Chenghao Xiao and Vaibhav Adlakha and Orion Weller and Siva Reddy and Niklas Muennighoff},
      year={2025},
      eprint={2502.13595},
      archivePrefix={arXiv},
      primaryClass={cs.CL},
      url={https://arxiv.org/abs/2502.13595}, 
}

@inproceedings{gte,
  title={mGTE: Generalized Long-Context Text Representation and Reranking Models for Multilingual Text Retrieval},
  author={Zhang, Xin and Zhang, Yanzhao and Long, Dingkun and Xie, Wen and Dai, Ziqi and Tang, Jialong and Lin, Huan and Yang, Baosong and Xie, Pengjun and Huang, Fei and others},
  booktitle={Proceedings of the 2024 Conference on Empirical Methods in Natural Language Processing: Industry Track},
  pages={1393--1412},
  year={2024}
}

@online{mxbaiv1,
  title={Boost Your Search With The Crispy Mixedbread Rerank Models},
  author={Aamir Shakir and Darius Koenig and Julius Lipp and Sean Lee},
  year={2024},
  url={https://www.mixedbread.ai/blog/mxbai-rerank-v1},
}

@misc{bge_reranker,
      title={C-Pack: Packaged Resources To Advance General Chinese Embedding}, 
      author={Shitao Xiao and Zheng Liu and Peitian Zhang and Niklas Muennighoff},
      year={2023},
      eprint={2309.07597},
      archivePrefix={arXiv},
      primaryClass={cs.CL}
}

@article{qwen3embedding,
  title={Qwen3 Embedding: Advancing Text Embedding and Reranking Through Foundation Models},
  author={Zhang, Yanzhao and Li, Mingxin and Long, Dingkun and Zhang, Xin and Lin, Huan and Yang, Baosong and Xie, Pengjun and Yang, An and Liu, Dayiheng and Lin, Junyang and Huang, Fei and Zhou, Jingren},
  journal={arXiv preprint arXiv:2506.05176},
  year={2025}
}

@misc{lamar,
      title={LAMAR: An Open Language-Aware Multilingual Alignment Reranker}, 
      author={Seongtae Hong and Youngjoon Jang and Jungseob Lee and Seungyoon Lee and Heuiseok Lim},
      year={2026},
      eprint={2607.22042},
      archivePrefix={arXiv},
      primaryClass={cs.IR},
      url={https://arxiv.org/abs/2607.22042}, 
}

@article{mxbaiv2,
  title={ProRank: Prompt Warmup via Reinforcement Learning for Small Language Models Reranking},
  author={Xianming Li and Aamir Shakir and Rui Huang and Julius Lipp and Benjamin Clavié and Jing Li},
  journal={arXiv preprint arXiv:2506.03487},
  year={2025}
}

@misc{lighton,
    title={One Adapter, Both Modalities: Field Notes from Building and Serving a Multimodal Reranker},
    author={Ananya, Ishrat Jahan and Chatelain, Amelie},
    year={2026},
    howpublished={\url{https://huggingface.co/blog/lightonai/lighton-rerank}},
}

@misc{zerank,
      title={zELO: ELO-inspired Training Method for Rerankers and Embedding Models}, 
      author={Nicholas Pipitone and Ghita Houir Alami and Advaith Avadhanam and Anton Kaminskyi and Ashley Khoo},
      year={2025},
      eprint={2509.12541},
      archivePrefix={arXiv},
      primaryClass={cs.AI},
      url={https://arxiv.org/abs/2509.12541}, 
}

@inproceedings{peng-etal-2025-efficiency,
    title = "Efficiency-Effectiveness Reranking {FLOP}s for {LLM}-based Rerankers",
    author = "Peng, Zhiyuan  and
      Wei, Ting-Ruen  and
      Song, Tingyu  and
      Zhao, Yilun",
    editor = "Potdar, Saloni  and
      Rojas-Barahona, Lina  and
      Montella, Sebastien",
    booktitle = "Proceedings of the 2025 Conference on Empirical Methods in Natural Language Processing: Industry Track",
    month = nov,
    year = "2025",
    address = "Suzhou (China)",
    publisher = "Association for Computational Linguistics",
    url = "https://aclanthology.org/2025.emnlp-industry.186/",
    doi = "10.18653/v1/2025.emnlp-industry.186",
    pages = "2782--2791",
    ISBN = "979-8-89176-333-3"
}

@misc{aarsen2026ettin-reranker,
    title = "Introducing the Ettin Reranker Family",
    author = "Aarsen, Tom",
    year = "2026",
    publisher = "Hugging Face",
    url = "https://huggingface.co/blog/ettin-reranker",
}

@inproceedings{qasper,
  title={A dataset of information-seeking questions and answers anchored in research papers},
  author={Dasigi, Pradeep and Lo, Kyle and Beltagy, Iz and Cohan, Arman and Smith, Noah A and Gardner, Matt},
  booktitle={Proceedings of the 2021 Conference of the North American Chapter of the Association for Computational Linguistics: Human Language Technologies},
  pages={4599--4610},
  year={2021}
}

@inproceedings{govreport,
  title={Efficient attentions for long document summarization},
  author={Huang, Luyang and Cao, Shuyang and Parulian, Nikolaus and Ji, Heng and Wang, Lu},
  booktitle={Proceedings of the 2021 conference of the north American chapter of the association for computational linguistics: Human language technologies},
  pages={1419--1436},
  year={2021}
}

@inproceedings{multinews,
  title={Multi-news: A large-scale multi-document summarization dataset and abstractive hierarchical model},
  author={Fabbri, Alexander Richard and Li, Irene and She, Tianwei and Li, Suyi and Radev, Dragomir},
  booktitle={Proceedings of the 57th annual meeting of the association for computational linguistics},
  pages={1074--1084},
  year={2019}
}

@inproceedings{bai2024longbench,
  title={Longbench: A bilingual, multitask benchmark for long context understanding},
  author={Bai, Yushi and Lv, Xin and Zhang, Jiajie and Lyu, Hongchang and Tang, Jiankai and Huang, Zhidian and Du, Zhengxiao and Liu, Xiao and Zeng, Aohan and Hou, Lei and others},
  booktitle={Proceedings of the 62nd annual meeting of the association for computational linguistics (volume 1: Long papers)},
  pages={3119--3137},
  year={2024}
}

@misc{gemmateam2026gemma4,
      title={Gemma 4 Technical Report}, 
      author={Gemma Team},
      year={2026},
      eprint={2607.02770},
      archivePrefix={arXiv},
      primaryClass={cs.CL},
      url={https://arxiv.org/abs/2607.02770}, 
}

@misc{qwen36_35b_a3b,
    title = {{Qwen3.6-35B-A3B}: Agentic Coding Power, Now Open to All},
    url = {https://qwen.ai/blog?id=qwen3.6-35b-a3b},
    author = {{Qwen Team}},
    month = {April},
    year = {2026}
}

@article{moreira2024nv,
  title={Nv-retriever: Improving text embedding models with effective hard-negative mining},
  author={Moreira, Gabriel de Souza P and Osmulski, Radek and Xu, Mengyao and Ak, Ronay and Schifferer, Benedikt and Oldridge, Even},
  journal={arXiv preprint arXiv:2407.15831},
  year={2024}
}

@inproceedings{li2025making,
  title={Making text embedders few-shot learners},
  author={Li, Chaofan and Qin, MingHao and Xiao, Shitao and Chen, Jianlyu and Luo, Kun and Lian, Defu and Shao, Yingxia and Liu, Zheng},
  booktitle={International Conference on Learning Representations},
  volume={2025},
  pages={38107--38124},
  year={2025}
}

@article{adamw,
  title={Decoupled weight decay regularization},
  author={Loshchilov, Ilya and Hutter, Frank},
  journal={arXiv preprint arXiv:1711.05101},
  year={2017}
}

@inproceedings{sbert,
    title = "Sentence-BERT: Sentence Embeddings using Siamese BERT-Networks",
    author = "Reimers, Nils and Gurevych, Iryna",
    booktitle = "Proceedings of the 2019 Conference on Empirical Methods in Natural Language Processing",
    month = "11",
    year = "2019",
    publisher = "Association for Computational Linguistics",
    url = "https://arxiv.org/abs/1908.10084",
}

@article{zhu2024longembed,
  title={LongEmbed: Extending Embedding Models for Long Context Retrieval},
  author={Zhu, Dawei and Wang, Liang and Yang, Nan and Song, Yifan and Wu, Wenhao and Wei, Furu and Li, Sujian},
  journal={arXiv preprint arXiv:2404.12096},
  year={2024}
}

@article{fan2026minireranker,
  title={miniReranker: Efficient Multimodal Reranking through Visual Cache Reuse and Interaction Sparsity},
  author={Fan, Yingqi and Lu, Xuan and Zhao, Anhao and Tong, Junlong and Nie, Ping and Zou, Kai and Ma, Yunpu and Zhang, Wei and Shen, Xiaoyu},
  journal={arXiv preprint arXiv:2606.10759},
  year={2026}
}
